\documentclass[11pt]{article}
\usepackage{graphicx}  
\usepackage{amsmath}  

\usepackage{fullpage}
\usepackage{authblk}   
\usepackage{setspace}  
\usepackage{textcomp}

\usepackage[cdot,mediumqspace,amssymb]{SIunits} 
\usepackage[autolanguage]{numprint} 
\usepackage{helvet}
\usepackage[sc]{mathpazo}
\usepackage[T1]{fontenc}

\usepackage[top=1in, bottom=1.12in, left=0.9in, right=0.9in]{geometry}

\renewcommand{\thefigure}{\arabic{figure}}
\renewcommand{\thetable}{\arabic{table}}
\renewcommand{\theequation}{\arabic{equation}}
\renewcommand{\thepage}{\arabic{page}}

\newcommand\startsupplement{%
    \makeatletter
       \setcounter{table}{0}
       \renewcommand{\thetable}{S\arabic{table}}
       \setcounter{figure}{0}
       \renewcommand{\thefigure}{S\arabic{figure}}
       \setcounter{equation}{0}
       \renewcommand{\theequation}{S\arabic{equation}}
       \setcounter{page}{1}
       \renewcommand{\thepage}{S\arabic{page}}
    \makeatother}
\usepackage{setspace}
\onehalfspace
\spacing{1}

\newenvironment{affiliations}{%
    \setcounter{enumi}{1}%
    \setlength{\parindent}{0in}%
    \slshape\sloppy%
    \begin{list}{\upshape$^{\arabic{enumi}}$}{%
        \usecounter{enumi}%
        \setlength{\leftmargin}{0in}%
        \setlength{\topsep}{0in}%
        \setlength{\labelsep}{0in}%
        \setlength{\labelwidth}{0in}%
        \setlength{\listparindent}{0in}%
        \setlength{\itemsep}{0ex}%
        \setlength{\parsep}{12pt}%
        }
    }{\end{list}\par\vspace{12pt}}

\makeatletter
\renewcommand{\maketitle}{\bgroup\setlength{\parindent}{0pt}
\begin{flushleft}
  \textbf{\@title}\\
  \vspace{12pt}
  \@author
\end{flushleft}\egroup
}
\makeatother

\makeatletter
\renewcommand\section{\@startsection{section}{1}{\z@}%
                                  {-3.5ex \@plus -1ex \@minus -.2ex}%
                                  {2.3ex \@plus.2ex}%
                                  {\large\bfseries}}

\renewcommand\subsection{\@startsection{section}{1}{\z@}%
                                  {-3.5ex \@plus -1ex \@minus -.2ex}%
                                  {2.3ex \@plus.2ex}%
                                  {\large\bfseries}}
\makeatother

\usepackage[font=rm, labelfont={rm,bf}]{caption}
\def\helvetica{hv}
\xdef\helvetica{\helvetica\space}
\def\frutigerbold{cmssbx10 }
=\frutigerbold at 24.5pt

\usepackage[sc]{mathpazo}
\usepackage[T1]{fontenc}
\usepackage{lettrine}

\title{\Large{\textbf{\center{All-optical control of a solid-state spin using coherent dark states}}}}

\author{\textsc{Christopher G. Yale}$^{a,1}$, \textsc{Bob B. Buckley}$^{a,1}$, \textsc{David J. Christle}$^a$, \textsc{Guido Burkard}$^b$, \textsc{F. Joseph Heremans}$^a$, \textsc{Lee C. Bassett}$^a$, and \textsc{David D. Awschalom}$^{a,2}$}

\begin{document}

\maketitle
\thispagestyle{empty}

\vspace{12pt}
\begin{affiliations}
 \item[$^a$ ] \small{Center for Spintronics and Quantum Computation, University of California, Santa Barbara, CA 93106, USA}
 \item[$^b$ ] \small{Department of Physics, University of Konstanz, D-78457 Konstanz, Germany}
\end{affiliations}
\noindent$^1$ These authors contributed equally\\
\noindent$^2$ e-mail: awsch@physics.ucsb.edu

\pagestyle{plain}
\setcounter{page}{1}
\pagenumbering{arabic}

\vspace{24pt}
\rm{
\textbf{The study of individual quantum systems in solids, for use as quantum bits (qubits) and probes of decoherence, requires protocols for their initialization, unitary manipulation, and readout. In many solid-state quantum systems, these operations rely on disparate techniques that can vary widely depending on the particular qubit structure. One such qubit, the nitrogen-vacancy (NV) center spin in diamond, can be initialized and read out through its special spin-selective intersystem crossing, while microwave electron spin resonance (ESR) techniques provide unitary spin rotations. Instead, we demonstrate an alternative, fully optical approach to these control protocols in an NV center that does not rely on its intersystem crossing. By tuning an NV center to an excited-state spin anticrossing at cryogenic temperatures, we use coherent population trapping and stimulated Raman techniques to realize initialization, readout, and unitary manipulation of a single spin. Each of these techniques can be performed directly along any arbitrarily-chosen quantum basis, removing the need for extra control steps to map the spin to and from a preferred basis. Combining these protocols, we perform measurements of the NV center's spin coherence, a demonstration of this full optical control. Consisting solely of optical pulses, these techniques enable control within a smaller footprint and within photonic networks. Likewise, this approach obviates the need for both ESR manipulation and spin addressability through the intersystem crossing. This method could therefore be applied to a wide range of potential solid-state qubits, including those which currently lack a means to be addressed.
}

\vspace{12pt}
\lettrine[lines=2, lraise=0.075, nindent=0em, slope=-.5em, loversize=0.07, lhang=0.1] {T}o explore control of individual quantum states, our experiments exploit coherent dark resonances that occur in a basic quantum mechanical level configuration known as a lambda $(\Lambda)$ system. This configuration, consisting of two lower energy states coherently coupled to a single excited state, has been observed in a wide array of systems including atoms \cite{Gray1978}, trapped ions, diamond nitrogen-vacancy (NV) centers \cite{Santori2006,Togan2011}, quantum dots \cite{Xu2008}, superconducting phase qubits \cite{Kelly2010}, and optomechanical resonators \cite{Dong2012}. In trapped ions, $\Lambda $ systems can additionally be exploited to drive stimulated Raman transitions providing unitary rotations of the qubit state \cite{Blinov2004,Wineland1998}. This versatile structure also forms the framework for a variety of other important advances in quantum science such as electromagnetically induced transparency \cite{Boller1991}, slow light \cite{Budker1999}, atomic clocks \cite{Vanier2005}, laser cooling \cite{Aspect1988}, and spin-photon entanglement \cite{Togan2010}.

Here, we use time-resolved methods and quantum state tomography to explore the dynamics of various optically driven processes within a solid-state $\Lambda$ system (Fig. \ref{fig:LambdaConfiguration}A). This allows us to demonstrate three all-optical quantum control \cite{Wineland1998,Hilser2012,Berezovsky2008} protocols for a single NV center: initialization, unitary rotation, and readout of its spin state. Our $\Lambda$ system consists of two ground state spin sublevels coupled to a spin-composite excited state sublevel formed by tuning the excited states to an avoided level anticrossing. Driving transitions between the levels of our $\Lambda$ system resonantly with appropriate coherent light fields (Fig. \ref{fig:LambdaConfiguration}A) causes any initial mixed state to be purified\cite{Murch2012}, or trapped, into a well-defined but selectable quantum superposition. This superposition is called the ``dark state'' since destructive interference from the driving fields causes the system not to be optically excited. This dissipative effect, known as coherent population trapping (CPT), allows us to initialize the precessing spin anywhere on the rotating-frame Bloch sphere, the geometric surface corresponding to all possible superposition states of the spin. Opposite the dark state on the Bloch sphere is a corresponding ``bright state'' which couples strongly to the optical fields. Together, these dark and bright states define a unique basis whose orientation within the rotating frame (SI Appendix) is a function of the relative phase and amplitude of the two driving optical fields (Fig. \ref{fig:LambdaConfiguration}A). A complementary process allows us to read out the spin state within this selected basis because the resultant photoluminescence (PL) during the transient period of the CPT interaction is proportional to the spin's projection along the bright state. Furthermore, detuning the driving fields from resonance within the $\Lambda$ system produces unitary rotations of the spin state about a chosen dark/bright-state axis, a dispersive technique that is a product of stimulated Raman transitions (SRT). Thus, this $\Lambda$ system approach allows spin initialization, readout, and rotation schemes to all function within a fully mutable basis.

\begin{figure}[!ht]
\begin{center}
\includegraphics[]{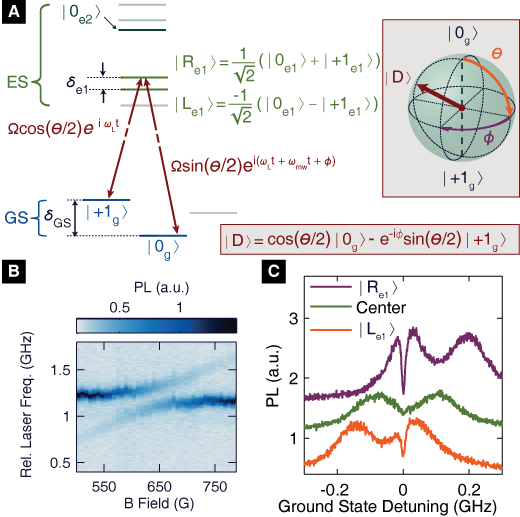}
  \caption[$\Lambda\ $ configuration and the NV center]{\label{fig:LambdaConfiguration}
\textbf{$\Lambda\ $ configuration and the NV center} \textbf{A}, $\Lambda\ $ configuration within the NV center level structure (left), depicting excitation with two optical driving fields from ground states (GS) to excited states (ES). At the center of the excited-state anticrossing, the two upper $\Lambda\ $ states $|R_{e1}\rangle$ and $|L_{e1}\rangle$ (bolded green) are the orthogonal, equal superpositions of $|0_{e1}\rangle$ and $|+1_{e1}\rangle$. An example dark state, $|D\rangle$, from the $|R_{e1}\rangle$ $\Lambda\ $ system, is plotted on the rotating-frame Bloch sphere (right), where its polar, $\theta$, and azimuthal, $\phi$, positions are a function of applied laser power and phase (equation). \textbf{B}, PL from resonant excitation as a function of magnetic field and laser frequency illustrating the anticrossing between the $|0_{e1}\rangle$ and $|+1_{e1}\rangle$. \textbf{C}, PL from excitation with two optical fields as a function of the detuning of $\omega_{mw}$ from $\delta_{GS}/\hbar$, resonant with either $|R_{e1}\rangle$, $|L_{e1}\rangle$, or centered between both resonances. }
\end{center}
\end{figure}

\section*{A $\Lambda$ System in the NV Center}

The negatively charged NV center consists of a substitutional nitrogen atom adjacent to a lattice vacancy within a diamond crystal\cite{Wrachtrup2006}. Its millisecond-scale coherence times\cite{Balasubramanian2009} are exceptional for a solid-state system, and coherent optical transitions enable important applications in quantum optics and quantum information processing such as spin-light coherence\cite{Buckley2010} and entanglement\cite{Togan2010}, single-shot readout\cite{Robledo2011}, coupling to photonic cavities\cite{Faraon2011a,Faraon2012}, two-photon interference\cite{Bernien2012a,Sipahigil2012}, implementations of quantum games\cite{George2013}, and photon-mediated spin-spin entanglement of distant NV centers\cite{Bernien2012b}. The ground-state spin triplet can be photoexcited both resonantly ($\sim$\unit{637}{\nano\meter}) and non-resonantly to an excited-state spin-triplet orbital doublet. We perform our measurements at cryogenic temperatures (\unit{8}{\kelvin}), where these excited-state levels become energetically narrow\cite{Fu2009} and their fine structure\cite{Batalov2009,Manson2006} can be tuned with magnetic, electric and strain fields\cite{Tamarat2006, Bassett2011}. The NV center's spin is traditionally addressed using its intersystem crossing\cite{Manson2006, Robledo2011a}, through which the spin is both polarized into the $m_s = 0$ spin sublevel under optical illumination and measured with spin-dependent PL intensity emitted in the NV center's phonon sideband (\unit{650-800}{\nano\meter}). While the unique attributes of the NV center's intersystem crossing have made its spin stand out as an optically-addressable qubit, the intersystem crossing is not necessary for our optical approaches to address the spin.

We select the subspace spanned by two ground-state spin-triplet sublevels ($m_s = 0$ and $+1$) as our qubit states; the presence of the third sublevel ($m_s = -1$) causes only a small loss in fidelity (SI Appendix). We denote these states $|0_g\rangle$and $|+1_g\rangle$. To form the necessary excited state, we apply a magnetic field to reach a spin sublevel anticrossing, whose levels are a function of crystal strain, spin-spin, spin-orbit, and Zeeman interactions\cite{Doherty2011, Maze2011} (Methods). The anticrossing we use is between the $|0_{e1}\rangle$ and $|+1_{e1}\rangle$ spin sublevels within the lower-energy excited-state orbital branch and results in two spin-composite levels (Fig. \ref{fig:LambdaConfiguration}B) separated in energy by $\delta_{e1}\sim h*$\unit{0.18}{\giga\hertz}. Either of these superposed levels, denoted $|R_{e1}\rangle$ and $|L_{e1}\rangle$, can act as the upper state of our $\Lambda$ system (Fig. \ref{fig:LambdaConfiguration}A).

In order to address the $\Lambda$-system transitions, we split light from a \unit{637}{\nano\meter} ($\omega_L/(2\pi)\sim$\unit{470,000}{\giga\hertz}) laser tunable across the NV center's optical transitions into sidebands (multiples of $\omega_{mw}/(2\pi)\sim$\unit{4.6}{\giga\hertz}) with an electro-optic phase modulator. The relative phase ($\phi$) between the two optical fields that are resonant with the $\Lambda$ transitions determines the azimuthal position of the dark state on the Bloch sphere in the $\omega_{mw}$ rotating frame. Similarly, the relative amplitude of the two optical fields determines the dark state's polar angle, $\theta$ (Fig. \ref{fig:LambdaConfiguration}A). We first observe CPT spectroscopically [1-6] by examining the PL under quasi-continuous photoexcitation that optically drives only one of the $\Lambda$ systems. A sharp dip in PL is observed centered at $\omega_{mw} = \delta_{GS}/\hbar$ where $\delta_{GS}$ is the mean energy splitting between the spin eigenstates (Fig. \ref{fig:LambdaConfiguration}C), indicating that the spin is being coherently trapped in the dark state. Because the spin-composite excited states are orthogonal in the $|0_{e1}\rangle$ and $|+1_{e1}\rangle$ spin subspace, the dark states from each of the separate $\Lambda$ systems have opposite azimuthal phases but the same polar position on the rotating-frame Bloch sphere for a given optical $\Lambda$-driving configuration. For this reason, when we tune the laser to equally excite both $\Lambda$ systems (``center'' curve in Fig. \ref{fig:LambdaConfiguration}C), their competing dark states quench the PL dip. For subsequent studies, we set $\omega_{mw} = \delta_{GS}/\hbar$ unless otherwise noted.

\section*{Coherent Population Trapping for Arbitrary-Basis Spin Initialization}
\begin{figure}[!hb]
\begin{center}
\includegraphics[]{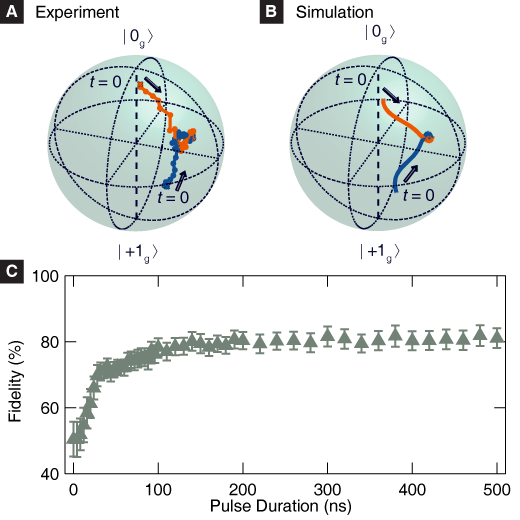}
  \caption[Time dynamics of coherent population trapping]{\label{fig:TimeDynamics}
\textbf{Time dynamics of coherent population trapping} \textbf{A}, Bloch sphere representation of the spin state as a function of the CPT interaction time, on resonance with $|R_{e1}\rangle$. Beginning near either $|0_g\rangle$ (orange) or $|+1_g\rangle$ (blue), this process polarizes the spin towards $|D\rangle$ regardless of its initial state. Errors are $\sim$3x the point size, and are detailed in the SI. \textbf{B}, Model of the time dynamics using a Lindblad master equation approach (description in SI). \textbf{C}, Fidelity of initialized spin state as a function of pulse duration. Fidelity is compared to the pure state $|D\rangle$. }
\end{center}
\end{figure}
We extend our investigation of the CPT interaction further by probing the time dynamics of the resultant spin state. We set the lasers resonant with the $|R_{e1}\rangle$ $\Lambda$ system to produce a dark state near the Bloch sphere equator. After preparing the initial spin state in either $|0_{g}\rangle$ or $|+1_{g}\rangle$ with traditional off-resonant (\unit{532}{\nano\meter} laser) optical polarization and microwave electron-spin resonance (ESR) techniques\cite{Jelezko2004}, we engage the CPT interaction for a variable duration to polarize the spin toward the dark state. We then perform quantum state tomography (Methods and SI Appendix) on the post-CPT spin state via microwave ESR pulses phase-matched to $\omega_{mw}$ and subsequent spin readout via a second laser resonant with the $|0_{g}\rangle$ to $|0_{e2}\rangle$ cycling transition\cite{Robledo2011} (Fig. \ref{fig:LambdaConfiguration}A and Methods). The tomographic reconstructions (Fig. \ref{fig:TimeDynamics}A) show that the spin state evolves towards the dark state regardless of its initial state, and a theoretical model accounting for both $\Lambda$ systems (Fig. \ref{fig:TimeDynamics}B) is in qualitative agreement with our data (Methods and SI Appendix). As a function of pulse duration, the initialization fidelity saturates at about 80\% after \unit{100}{\nano\second} (Fig. \ref{fig:TimeDynamics}C). The fidelity is limited, in particular, by decoherence from optical coupling to the second nearby $\Lambda$ system, as well as finite $T_2^*$ spin coherence\cite{Dobrovitski2008}, spectral diffusion\cite{Fu2009}, and some pumping into the third spin state $|-1_g\rangle$ (SI Appendix).

\begin{figure}[!hb]
\begin{center}
\includegraphics[]{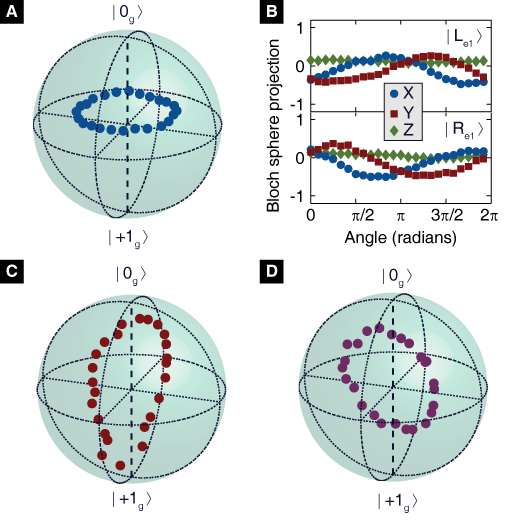}
  \caption[Arbitrary spin-state initialization]{\label{fig:ArbInit}
\textbf{Arbitrary spin-state initialization} \textbf{A}, Azimuthal initialization of spins via CPT on resonance with $|R_{e1}\rangle$. Varying the relative phase between the two optical fields ($\phi$) changes the azimuthal location. \textbf{B}, X, Y, and Z projections of azimuthal initialization, on resonance with $|L_{e1}\rangle$ (top) vs. the orthogonal state $|R_{e1}\rangle$ (bottom). Error bars are within point size.  \textbf{C}, Polar initialization of spins, on resonance with $|R_{e1}\rangle$. Varying the relative amplitude between the two optical fields ($\tan(\theta/2)$) changes the polar location. \textbf{D}, Initialization of spins along a great circle canted $\pi/4$ off the polar axis, achieved through control of both the relative phase and amplitude of the two optical fields. Prior to CPT, the spin was polarized into $|+1_{g}\rangle$ (SI). The CPT pulse duration was \unit{200}{\nano\second} (\textbf{A}, \textbf{C}, \textbf{D}) or \unit{100}{\nano\second} (\textbf{B}). Errors are $\sim$2x the point size, and are detailed in the SI. }
\end{center}
\end{figure}
The allure of this technique is the ability to initialize the spin arbitrarily on the Bloch sphere solely by varying the relative phase and amplitude of the two optical fields. In Fig. \ref{fig:ArbInit}A, we demonstrate initialization along different equatorial points of the Bloch sphere by changing the relative phase between the two driving optical fields resonant with $|R_{e1}\rangle$. Because $|R_{e1}\rangle$ and $|L_{e1}\rangle$ are orthogonal spin mixtures, tuning the lasers to $|L_{e1}\rangle$ instead is equivalent to shifting $\phi$ of the final state by $\pi$ radians (Fig. \ref{fig:ArbInit}B). Alternatively, by tuning the relative amplitudes of the two optical fields, we initialize the spin at various points along a meridian of the Bloch sphere (Fig. \ref{fig:ArbInit}C). Finally, we combine polar and azimuthal control to demonstrate spin initialization at points along a great circle rotated $\pi$/4 radians from the polar axis (Fig. \ref{fig:ArbInit}D).

\section*{Arbitrary-Basis Spin Readout via CPT Photoexcitation}
\begin{figure}[!h]
\begin{center}
\includegraphics[]{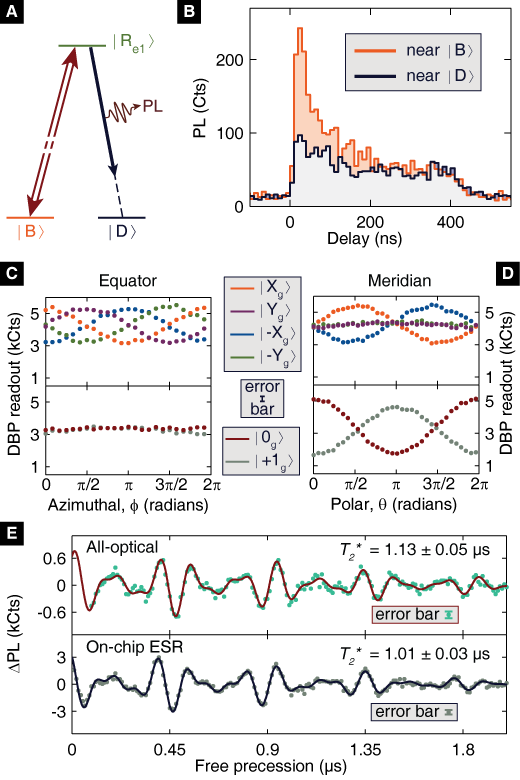}
  \caption[Arbitrary-basis spin-state readout]{\label{fig:ArbReadout}
\textbf{Arbitrary-basis spin-state readout} \textbf{A}, $\Lambda$ configuration recast in terms of ground state orthogonal superpositions, the bright $|B\rangle$ and dark $|D\rangle$ states. The driving fields are similarly recast as an optical pump on the bright state transition. \textbf{B}, The emitted PL response of the NV center spin as it settles into the dark state, starting either near the bright or dark state. This trace is a sum of $2.3 \times 10^6$ iterations with the data binned into \unit{10}{\nano\second} time intervals. \textbf{C}, Spins initialized at points along the equator and read out through DBP. The DBP basis is chosen such that the corresponding bright state, indicated in the legend, is at one of four points on the equator (top panel) or one of the poles (bottom panel). \textbf{D}, Spins initialized at points along a meridian, mapping out Rabi oscillations, and read out via DBP in the same bases as in C. \textbf{E}, (top) All-optical Ramsey experiment, detuned such that $\omega_{mw} - \delta_{GS}/\hbar = 2\pi\cdot$\unit{7.5}{\mega\hertz}. The CPT initialization and DBP readout pulses are each \unit{50}{\nano\second} in duration. (bottom) Room-temperature Ramsey measurement using ESR pulses with similar detuning for comparison. All error bars represent $1\sigma$ shot noise.}
\end{center}
\end{figure}
Readout along an arbitrarily-chosen basis\cite{Kosaka2009} is realized through a complementary process as the emitted PL during the CPT interaction with the two optical fields is proportional to the projection of the spin along the bright-state axis. This can be thought of as a recasting of the ground states of our $\Lambda$~system in terms of the bright and dark states, orthogonal superpositions of the original spin eigenstates $|0_g\rangle$ and $|+1_g\rangle$. The two driving light fields are correspondingly recast as a single optical pump acting on the bright state transition (Fig. \ref{fig:ArbReadout}A), since they do not couple to the dark state from destructive interference of photoexcitation. During the interaction, the spin evolves toward the dark state, and the emitted PL provides a measure of the spin state prior to the interaction (Fig. \ref{fig:ArbReadout}B). This technique, which we refer to as dark/bright-state projection (DBP), bears similarity to electromagnetically induced transparency\cite{Boller1991}, but we instead measure the transient optical response of the NV center rather than the amount of transmitted light.

To demonstrate arbitrary-basis readout, we prepare the spin state with ESR pulses either along various positions on either the equator (Fig. \ref{fig:ArbReadout}C) or a meridian (Fig. \ref{fig:ArbReadout}D) of the Bloch sphere, and then use DBP to read out the spin state along six separate bases with bright states corresponding to the $\pm X$, $\pm Y$, and $\pm Z$ positions of the rotating frame Bloch sphere. The number of photons measured is in direct proportion to the projection of the spin state along the chosen axis. The signal-to-noise of spin readout using DBP along polar bright states is comparable to traditional spin readout techniques via the intersystem crossing, while DBP spin readout along equatorial states requires roughly 3x more averaging (SI Appendix) to achieve a similar signal-to-noise ratio. By combining both CPT initialization and DBP readout, we perform an all-optical Ramsey measurement\cite{Thomas1982} by varying the delay between the CPT and DBP pulses in order to measure the transverse inhomogeneous spin coherence time, $T_2^*$ (Fig. \ref{fig:ArbReadout}E, top). Collapses and revivals in the signal are indicative of hyperfine coupling to the $^{14}N$ spin. The all-optical response is similar to Ramsey measurements taken at room temperature using ESR pulses and traditional intersystem crossing-based initialization and readout (Fig. \ref{fig:ArbReadout}E, bottom).

\section*{Arbitrary-Axis Spin Rotations via Stimulated Raman Transitions}

Within this same optical coupling framework, we also demonstrate unitary spin rotations about any qubit axis via SRT. By detuning $\omega_L$ from resonance while keeping $\omega_{mw} = \delta_{GS}/\hbar$, driving the $\Lambda$~system produces adiabatic energy shifts of the bright state during the laser pulse without modifying the dark state energy, generating unitary spin rotations\cite{Blinov2004,Wineland1998,Berezovsky2008,Buckley2010} along the dark/bright state Bloch sphere axis. In order to drive rotations about an equatorial axis, we tune the two equal-intensity ($\tan(\theta/2)=1$) driving fields to be centered between the $|R_{e1}\rangle$ and $|L_{e1}\rangle$ resonances, such that SRT generated from both $\Lambda$ systems add constructively while CPT effects from both reduce coherence but produce no net spin polarization due to competing dark states. In Fig. \ref{fig:AllOptControl}A, we present the dynamics of SRT spin rotations along two different equatorial rotation axes of the Bloch sphere (``$\sigma_X$'' or ``$\sigma_Y$''), corresponding to different relative phases ($\phi$) of the two optical fields. We measure a 69\% process fidelity for a ``$\sigma_X$'' or ``$\sigma_Y$'' $\pi$-rotation, limited largely by spontaneous decay. Rotations about non-equatorial axes, such as the polar axis\cite{Buckley2010} (``$\sigma_Z$''), are also achievable in this system (Fig. \ref{fig:AllOptControl}B) but require different configurations of the light fields (SI Appendix).

\begin{figure}[!ht]
\begin{center}
\includegraphics[width=88 mm]{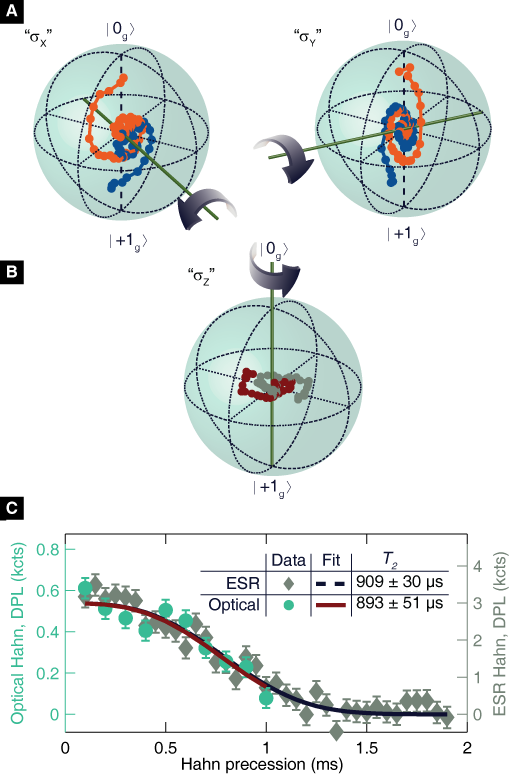}
  \caption[All-optical control of the NV center spin]{\label{fig:AllOptControl}
\textbf{All-optical control of the NV center spin} \textbf{A}, Bloch sphere representation of ``$\sigma_X$'' and ``$\sigma_Y$'' coherent rotations at $\sim$\unit{10}{\mega\hertz} due to SRT. The two measurements correspond to different relative EOM driving phases ($\phi$), separated by $\pi/2$ radians, and show the trajectory of a spin originating near $|0_g\rangle$. (orange) and $|+1_g\rangle$. (blue). The axes of rotation are added as guides to the eye. \textbf{B}, Bloch sphere representation of ``$\sigma_Z$'' coherent rotations due to SRT, showing spin trajectories originating near orthogonal points on the equator (maroon and grey). For \textbf{A}. \& \textbf{B}., errors are $\sim$2x the point size, and are detailed in the SI. \textbf{C}, All-optical Hahn echo measurement (green points) consisting of CPT spin initialization, a $\pi$ spin rotation via SRT, and DBP spin readout fit according to the equation on the graph. Room temperature Hahn measurement via ESR pulses is also shown (grey points). Error bars are $1\sigma$ shot noise.}
\end{center}
\end{figure}

Finally, to illustrate the full suite of these optical control protocols, we present an all-optical Hahn echo measurement of an NV center spin's homogeneous spin coherence time, $T_2$. This measurement consists of a CPT laser pulse for spin initialization along the Bloch equator, followed by a SRT laser pulse to flip the spin to produce an echo, and finally a DBP readout pulse to measure the final spin state along an equatorial basis (SI Appendix). We determine $T_2 \sim$\unit{900}{\micro\second}, corroborated by an ESR-based Hahn echo measurement at room temperature (Fig. \ref{fig:AllOptControl}C).

\section*{Conclusions}
We demonstrate all-optical initialization, readout, and coherent unitary rotations of an individual NV-center spin, forming a triumvirate of protocols for single-spin control that can be performed along any arbitrarily-chosen basis. Using these protocols, we demonstrate two measurements of transverse spin coherence solely with optical pulses. The ability to select any basis allows for quantum operations to be implemented directly without the need for extra control steps to project onto or from the preferred energy eigenstate basis. This eliminates the need for ESR operations\cite{Jelezko2004}, enabling control of individual spins within a much smaller device footprint, with promise for large-scale implementations of spin arrays\cite{Toyli2010} or photonic networks\cite{Faraon2011a,Faraon2012}. Perhaps most importantly, these methodologies mitigate the need for the NV center's intersystem crossing spin-selectivity and thus can be used to investigate and control a wide array of defects and other localized quantum states in solid-state materials, not just those with NV-like structures\cite{Weber2010,Koehl2011}. As such, these techniques open the door to exploring quantum coherence and developing quantum information platforms in a broad range of semiconductors and nanostructures.

\section*{Methods}

\textsc{Sample - }The sample was a $2\times2\times 0.5$\nobreakspace mm electronic grade diamond purchased from Element Six, consisting of $<5~\rm{ppb}$ nitrogen that was irradiated with a $1\rm{e}14~\rm{electrons/cm^2}$, $2~\rm{MeV}$ dose and subsequently annealed at $850~\degree$C for two hours. Ti/Pt/Au devices, consisting of DC pads and a short-terminated waveguide, were deposited on the sample using standard photolithographic techniques. All experiments were performed in a confocal microscopy setup (SI Appendix) with a liquid helium flow cryostat held at \unit{8}{\kelvin}. The sample was thermally sunk to the cryostat and the waveguide was wirebonded to a microwave line in the cryostat for on-chip ESR. The studied NV center excited state orbital strain splitting between $m_s=0$ spin sublevels varied from \unit{4.6}{\giga\hertz} to \unit{5.8}{\giga\hertz} between cryostat cooldowns as thermal cycling modified the crystal strain. As a result, the DC-applied magnetic field at which the lower-branch excited state spin anticrossing occurred varied ($550-750~\rm{G}$) which led to variations in the ground state spin splitting between $|0_g\rangle$ and $|+1_g\rangle$ ($\delta_{GS}/h \sim$\unit{4.3}{} to \unit{5.0}{\giga\hertz}).

\vspace{12pt}
\noindent\textsc{Experimental Techniques - }Our confocal microscopy setup consists of a \unit{100}{\milli\watt} \unit{532}{\nano\meter} non-resonant excitation laser used for the standard NV center intersystem crossing spin initialization and readout protocol\cite{Wrachtrup2006}, and two tunable \unit{637}{\nano\meter} lasers resonant with various NV center optical transitions. The light field from one of the resonant lasers was fiber-coupled to an EOM in order to split the optical field, at $\omega_L$, into different frequency sidebands, separated by $\omega_{mw}$, to optically drive the $\Lambda$ system. For the CPT and SRT measurements, a second resonant laser functioned as a one-color spin-state readout laser along the $S_Z$ basis by being resonant with the $|0_g\rangle$ to $|0_{e2}\rangle$. transition\cite{Robledo2011} (Fig. \ref{fig:LambdaConfiguration}A) resulting in higher collected PL when the spin was in $|0_g\rangle$. In Fig. \ref{fig:AllOptControl}c only, the light field from this second resonant laser was instead fiber-coupled to a second EOM, where the first laser was used to perform CPT and DBP and the second laser was used to perform SRT for this Hahn echo pulse sequence. All three lasers were gated using separate acousto-optic modulators (AOMs) for pulse timing control. They were subsequently passed through a variety of polarization optics, combined with beamsplitters, and focused onto the sample with a $0.85$\nobreakspace numerical aperture $100\times$ microscope objective that is aberration-corrected for the cryostat window. PL from the NV's red-shifted phonon sideband was collected back through the objective, filtered by dichroic beamsplitters, and focused onto a silicon avalanche photodiode (APD).

Microwaves to drive the EOM(s) and for on-chip microwave ESR driving\cite{Jelezko2004} originated from the same signal generator at frequency $\omega_{mw}/(2\pi)$, which varied from \unit{4.3 - 5.0}{\giga\hertz} due to variations in $\delta_{GS}/h$ from changes in NV center strain. The microwaves going to the EOMs and to the sample for ESR passed through IQ modulators for phase control between the various CPT, SRT, DBP, and ESR pulses. These microwave signals were also gated in time and amplitude-controlled using microwave modulators and switches. Timing for the microwave switches, AOMs, and IQ modulators were controlled by an arbitrary waveform generator, a PulseBlaster card, and a pulse-pattern generator. Pulse sequences used for these experiments consisted of a traditional initialization pulse at \unit{532}{\nano\meter} as a spin reset, followed by a sequence consisting of a number of the following techniques: on-chip microwave ESR pulses as well as techniques utilizing the \unit{637}{\nano\meter} tunable lasers, including CPT spin-state initialization, DBP spin-state readout, $|0_{e2}\rangle$ spin-state readout, and/or SRT coherent spin rotation. Details for each pulse sequence are discussed in the SI Appendix. A magnetic field was applied along the NV center axis with a permanent magnet on a motorized stage and was adjusted to tune to the anticrossing used.

\vspace{12pt}
\noindent\textsc{Quantum State Tomography - }To perform quantum state tomography on our CPT spin-state initialization and SRT coherent rotation, we read out the $X$, $Y$, and $Z$ projections of the post-interaction state. All projections were mapped onto the $S_Z$ basis using ESR pulses and then read out with the laser resonant with $|0_{e2}\rangle$. We applied a Bayesian approach to the tomographical reconstruction of the spin state\cite{BlumeKohout2010}, detailed in the SI Appendix, that takes into account finite readout contrast, laser drift, and axial/length imperfections in the microwave rotations used to project the different spin components.

\vspace{12pt}
\noindent\textsc{Theoretical Modeling - }To describe the dynamics of the NV center spin under optical excitation in the $\Lambda$ level configuration\cite{Fleischhauer2005}, we include five energy levels: two out of the three ground-state levels $|0_g\rangle$, $|+1_g\rangle$, the two mixed excited states $|L_{e1}\rangle$ and $|R_{e1}\rangle$ as the upper state of each $\Lambda$ system, as well as the intermediate singlet $|S\rangle$ which here plays a role for unintentional intersystem crossings. The Hamiltonian, in the rotating frame, for the subspace spanned by these five basis states can be expressed as

\begin{equation}
H = \sum_\alpha \epsilon_\alpha |\alpha\rangle\langle\alpha|
     + \sum_{G=0,1}\sum_{E=R,L}\Big(\Omega_{GE} |E_{e1}\rangle\langle G_g| + h.c.\Big),
\end{equation}

\noindent where the first sum runs over all states $\alpha = 0_g, +1_g, L_{e1}, R_{e1}, S$ with
corresponding energies
${\epsilon_{0_g} = \epsilon_{+1_g} = \Delta_{L}}$ (where $\Delta_L$ is detuning of $\omega_L$  from resonance to a $\Lambda$ system), ${\epsilon_{R_{e1}} = 0}$, ${\epsilon_{L_{e1}} = -\delta_{e1}}$,
and $\epsilon_S$.
The laser excitation from one of the lower
states ${G=0,1}$ to one of the upper states ${E=L,R}$ is described by the Rabi frequencies
in the rotating frame,
\begin{eqnarray}
\Omega_{1E} = \Omega \cos(\theta/2) \\
\Omega_{0E} = \pm \Omega \sin(\theta/2) e^{i \phi}
\end{eqnarray}
where the upper (lower) sign holds for ${E=R}$ (${E=L}$).

We studied the time evolution of the system by numerically solving the Lindblad master equation\cite{Lindblad1976,Breuer2002} for the density matrix of the NV center in the rotating frame. In addition to coherent processes such as excitation from the two driving fields, the master equation also accounts for spontaneous decays of charge and spin with some rates known from independent experiments. In the idealized, long-time limit case, with only one excited level included, the resulting eigenvector with eigenvalue $0$ corresponds to the dark state:

\begin{equation}
|D\rangle = \cos(\theta/2)|0_g\rangle - e^{\mp i \phi} \sin(\theta/2)|+1_g\rangle
\end{equation}

\noindent where the upper (lower) sign holds for the single excited state level being $E = R (E = L)$. In actuality, the steady state is described by a mixed state which can deviate slightly from $|D\rangle\langle D|$. The simulated behavior of the NV spin during CPT and SRT is in good qualitative agreement with the experimental data (Fig. \ref{fig:TimeDynamics} and Fig. \ref{fig:InitModel}-\ref{fig:ZRot}). Further details can be found in the SI Appendix.

\section*{Acknowledgments}
We thank J. Bochmann, G. Calusine, A. L. Falk, and D. M. Toyli for discussions and K. Ohno for sample preparation. This work was supported by the Air Force Office of Scientific Research, the Army Research Office, and the Defense Advanced Research Projects Agency. G.B. acknowledges funding from DFG within SFB767, from the Konstanz Center for Applied Photonics (CAP), from BMBF QuHLRep, and from the Research Initiative UltraQuantum.


\section*{Author Contributions}
C.G.Y., B.B.B., G.B., and D.D.A. designed the experiment. C.G.Y. and B.B.B. performed the measurements. D.J.C. performed the tomographic reconstruction analysis. G.B. developed the theoretical modeling. C.G.Y., F.J.H., B.B.B., and L.C.B. designed and fabricated the sample. F.J.H., C.G.Y., and B.B.B. developed the figures. All authors analyzed the experiment and contributed to writing the paper.


\clearpage
\startsupplement
\LARGE{\textbf{Supporting Information:}}

\fontsize{11pt}{12}\selectfont
\maketitle
\tableofcontents
\vspace{24pt}
\section{Experimental details}

\vspace{12pt}

\lettrine[lines=2, lraise=0.075, nindent=0em, slope=.5em, loversize=0.07, lhang=0.1] {A} schematic of our confocal setup is provided in Fig. \ref{fig:ExpSetup} that incorporates a continuous wave (CW) green diode laser (\unit{532}{\nano\meter}), two CW tuneable red (\unit{637}{\nano\meter}) diode lasers with optional sideband wavelengths generated via electro-optic phase modulators (EOMs), NV center photoluminescence (PL) collection (\unit{650-800}{\nano\meter}) via an avalanche photodiode, and on-chip microwave electron spin resonance techniques (ESR) all gatable in time.  All timing sequences used were programmed into an arbitrary waveform generator and are described later in this Supplementary Information.

While the effects we presented in the main text were observed in multiple NV centers, care was taken to select an NV center that  had several desirable properties.  We chose an NV center with a reasonable $T_2^*$ time ($\sim$\unit{1}{\micro\second}) due to the finite duration of the CPT/SRT/DBP pulses.  Because the energy separation between $|R_{e1}\rangle$ and $|L_{e1}\rangle$ $\Lambda$ systems was quite small ($\delta_{e1}/h=$ \unit{0.18}{\giga\hertz}), it was desirable to choose an NV center with a narrow inhomogeneously-broadened optical linewidth (\unit{0.05}{\giga\hertz} FWHM for the NV center presented).  We also chose an NV center that had a transverse strain splitting which varied between \unit{4.6}{\giga\hertz} and \unit{5.8}{\giga\hertz} as measured between the two orbital $m_s = 0$ spin sublevels.  The variations in this NV center's strain occurred between cooldowns due to thermal cycling of the cryostat.  The strain was in a range which allowed us to produce an avoided level crossing (anticrossing) in the lower excited state orbital branch between the $m_s = 0$ and $m_s = +1$ spin sublevels with the application of an external magnetic field along the N-V axis, ($550-750$ Gauss).  The resulting ground state spin splitting at these magnetic fields was relatively high ($\delta_{GS} / h =$ \unit{4.3-5}{\giga\hertz}) allowing for the EOM sidebands to be widely spaced, meaning the unused laser harmonics generated by the EOM were far from any NV center resonances.

\begin{figure}[h]
\begin{center}
\includegraphics[width = 172 mm]{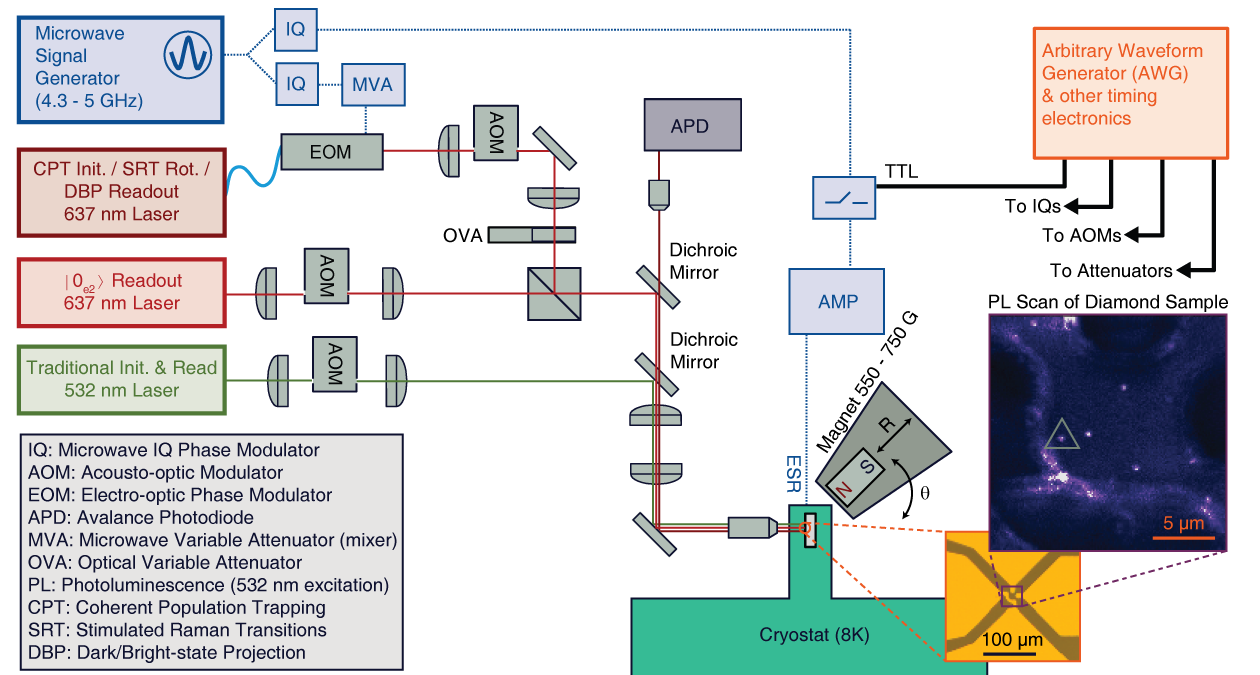}
  \caption[Schematic of experimental setup.]{\label{fig:ExpSetup}
\textbf{Schematic of experimental setup.} Diagram detailing the optical excitation paths and photon collection path along with the microwave electronics, timing electronics, cryostat, and magnetic field as described in Methods.  A PL scan of the region 6\nobreakspace\unit{}{\micro}$\sf{m}$ below diamond surface, where the NV centre investigated is located (within the smoke-colored triangle).  A short-terminated on-chip waveguide wire used to apply microwave ESR pulses for ground state spin manipulation is visible in the lower left.  Deposited metallic pads on the right and top of the image are for applying dc voltages to the sample to affect the orbital splitting of NV center if necessary, but were not used in the present experiment. }
\end{center}
\end{figure}

It is also possible to generate an anticrossing in the upper excited state orbital branch but the resulting anticrossed eigenstates are much closer together in energy, making it prohibitively difficult to couple to an individual resonance.  Off-axis magnetic fields could be used to increase the splitting between anticrossed levels, but they would also reduce the spin-selectivity of the intersystem crossing (ISC) used for comparison to the CPT-based readout.

In order to fully control the phase and amplitude of the CPT laser system, we adjust properties of the microwaves driving the  EOM.  To adjust the dynamic phase between the two used light fields, we control the phase of the $\omega_{mw}$ microwaves driving the EOM with an IQ modulator, which moves the dark state azimuthally about the rotating-frame Bloch sphere.  We adjust the relative amplitude of the two used laser driving fields by varying the microwave power driving the EOM using a mixer, allowing us to adjust the polar position of the dark state on the Bloch sphere.  While the relative amplitude of the two resonant optical fields adjusts as expected, the summed amplitude of the two $\Lambda$-resonant sidebands also changes.  This is because the EOM splits light at $\omega_L$ into several harmonics (of which we only select two) separated by $\omega_{mw}$, with amplitudes of the harmonics being a function of the amplitude of EOM microwave driving power in the form of a Bessel function.  To correct for this, we use an optical variable attenuator to compensate for this overall amplitude variance to within an order of magnitude in an attempt to keep $\Omega$ (Fig. 1A) fixed.  In addition, the overall frequency of the tunable laser, $\omega_L$, would drift, and due to the small separation in the anticrossing between $|R_{e1}\rangle$ and $|L_{e1}\rangle$, $\delta_{e1} / h = $ \unit{0.18}{\giga\hertz}, we needed to occasionally recenter the laser frequency to the appropriate tuning on the order of every 10 minutes with \unit{0.02}{\giga\hertz} laser frequency resolution.

\vspace{24pt}
\section{Quantum state tomography of arbitrary initialization and rotation}
\vspace{12pt}

In order to analyze our initialization and control protocols, we performed Bayesian quantum state tomography to characterize the various process output states and compute the corresponding fidelities found in the main text. This approach allows for an accurate statistical inversion of repeated projective measurements that are subject to both stochastic and systematic error. In contrast to maximum likelihood estimation, this approach always yields both point estimates and corresponding error bars that are physical for states near the boundaries of the allowed state space; moreover, it relaxes the assumption of asymptotic normality, achieving consistent estimates in the face of a finite number of measurements \cite{BlumeKohout2010}.

\begin{figure}[h!]
\begin{center}
\includegraphics{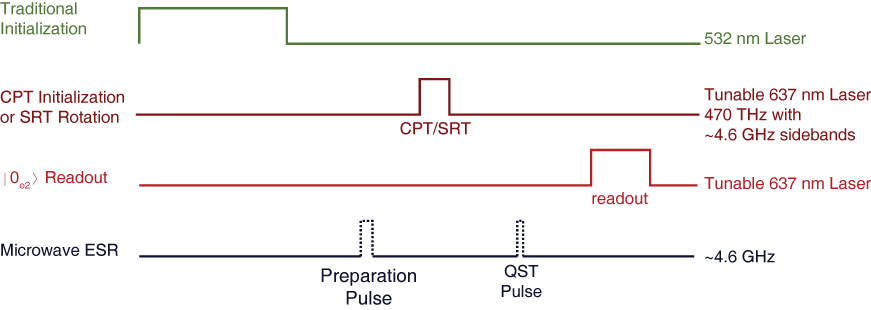}
  \caption[]{\label{fig:QSTPulse}
\textbf{Pulse sequence for arbitrary initialization and rotation.} The above pulse sequence was used for the data presented in Main Text Fig. 2, Fig. 3, and Fig. 5A,B.  All experiments investigating these protocols consisted of $\sf{\sim\!10^6}$ iterations of this pulse sequence to achieve a sufficient signal-to-noise ratio. }
\end{center}
\end{figure}

To individually study CPT or SRT, the NV center spin state is first prepared using non-resonant \unit{532}{\nano\meter} laser light to both mitigate photoionization and polarize the NV center ground state spin into $m_s = 0$ via ISC decay with roughly 75-80$\%$ fidelity \cite{Robledo2011a,Toyli2012}.  The state is then prepared on various places of the Bloch sphere with an ESR ``preparation pulse'' before either a CPT or SRT red laser pulse polarizes or rotates the spin, respectively. After the red laser, an additional ESR ``QST'' pulse is applied to rotate the X, Y, or Z spin projection onto the $|0_{g}\rangle / |\!+\!1_{g}\rangle$ readout measurement basis, phase-synced with the microwaves driving the EOM.  The timing of this QST pulse is chosen to coincide with the constructive rephasing of the three $^{14}\mathrm{N}$ hyperfine Larmor frequencies, corresponding to a delay of \unit{450}{\nano\second}.  Alternatively, a $\pi$-pulse can be added before the QST pulse to induce an echo of the spin coherence rather than use this rephasing.  Finally, the spin state is read out along this basis by measuring PL intensity during a single-color red laser resonant with the $|0_{g}\rangle / |0_{e2}\rangle$ cycling optical transition \cite{Robledo2011a}.

As a normalization, the state is more fully initialized into $|0_{g}\rangle$ (or $|\!+\!1_{g}\rangle$) after \unit{532}{\nano\meter} excitation by subjecting the spin to \unit{637}{\nano\meter} laser light for \unit{1}{\micro\second} resonant with the opposite $|\!+\!1_{g}\rangle$ (or $|0_{g}\rangle$) spin state and one of the $\Lambda$ systems.  This scheme depletes the aforementioned optically-driven sublevel in the ground state and populates its counterpart $|0_{g}\rangle$ (or $|\!+\!1_{g}\rangle$).  The readout contrast, $\mathcal{C}$, between the $|0_{g}\rangle$ and $|\!+\!1_{g}\rangle$ sublevels increases from about 62$\%$ from ISC spin polarization alone to about 84$\%$ after this purification. Because none of the CPT (or SRT) interactions appeared in practice to polarize the state beyond this level, this purified contrast serves as a consistent normalization for each state reconstruction and fidelity.

To begin our analysis, in terms of the density matrix, $\hat{\rho}$, the expectations are
\begin{align*}
\langle \mathrm{X} \rangle & = \mathrm{Tr}\left(\sigma_z U_{\mathrm{Y}_{-\!\frac{\pi}{2}}} \hat{\rho} U_{\mathrm{Y}_{-\!\frac{\pi}{2}}}^{\dag} \right) \\
\langle \mathrm{Y} \rangle & = \mathrm{Tr}\left(\sigma_z U_{\mathrm{X}_{\frac{\pi}{2}}} \rho U_{\mathrm{X}_{\frac{\pi}{2}}}^{\dag} \right) \\
\langle \mathrm{Z} \rangle & = \mathrm{Tr}\left(\sigma_z \rho  \right),
\end{align*}
while the expected fluorescence levels (in photon counts) are defined by the resonant laser normalizations,
\begin{align*}
\langle \mathcal{F}_{\langle \mathrm{X} \rangle} \rangle & = \mathcal{F}_{|0_{g}\rangle}  \left(1 - \frac{\mathcal{C}}{2}\right) + \mathcal{F}_{|0_{g}\rangle} \frac{\mathcal{C}}{2} \langle \mathrm{X} \rangle\\
\langle \mathcal{F}_{\langle \mathrm{Y} \rangle} \rangle & = \mathcal{F}_{|0_{g}\rangle}  \left(1 - \frac{\mathcal{C}}{2}\right) + \mathcal{F}_{|0_{g}\rangle} \frac{\mathcal{C}}{2} \langle \mathrm{Y} \rangle\\
\langle \mathcal{F}_{\langle \mathrm{Z} \rangle} \rangle & = \mathcal{F}_{|0_{g}\rangle}  \left(1 - \frac{\mathcal{C}}{2}\right) + \mathcal{F}_{|0_{g}\rangle} \frac{\mathcal{C}}{2} \langle \mathrm{Z} \rangle.
\end{align*}

We treat each of the data, $D_k$, as subject to normal error $\sigma_k$ from the model prediction $\langle F \rangle_k$ whose other parameters ($\mathcal{F}_{|0_{g}\rangle}$, $\mathcal{C}$, and those described below) are contained in a vector $\mathbf{X}$ such that
\begin{equation}
\mathrm{prob}\left(\mathbf{D} | \hat{\rho}, \sigma_k, \mathbf{X}\right) = \prod_{k} \frac{1}{\sqrt{2 \pi} \sigma_k} \mathrm{exp}\left(-\frac{\left(D_{k} - \langle F \rangle_k\right)^2}{2\sigma_k^2}\right).
\end{equation}
Because of shot noise, the lower limit to $\sigma_k$ is about $\sqrt{\langle F \rangle_k}$ but due to both the drift of the laser and stage mechanics, $\sigma_k$ was about a factor of two to five larger in practice. To capture this uncertainty in the expected mismatch, we set
\begin{equation}
\mathrm{prob}\left(\sigma_k | \overline{\sigma}_k \right) = \frac{2 \overline{\sigma}_k}{\sqrt{\pi} \sigma_k^2} \mathrm{exp}\left(-\frac{\overline{\sigma}_k^2}{\sigma_k^2}\right),
\end{equation}
so that we marginalize each $\sigma_k$ around a region of order $\overline{\sigma}_k$, expressing that $\sigma_k$ should be on the order of, but not necessarily equal to, $\overline{\sigma}_k$; this sort of construction makes our estimation statistically robust against data of unusually large drift (outliers). After performing the integration over all positive $\sigma_k$, we have that
\begin{equation}
\mathrm{prob}\left(\mathbf{D} | \hat{\rho}, \bar{\sigma}_k, \mathbf{X}\right) = \left( \sqrt{2} \pi \overline{\sigma}_k \left( 1 + \frac{\left(D_{k} - \langle F \rangle_k\right)^2}{2 \overline{\sigma}_k^2} \right) \right)^{-1}.
\end{equation}
In our analysis, we set $\bar{\sigma}_k = 2 \sqrt{\langle F \rangle_k}$. This likelihood is also used for normalization parameters $\mathcal{F}_{|0_{g}\rangle}$ and $\mathcal{C}$ in the posterior probability density to infer them simultaneously with the axial projections and systematic errors.

In addition to the random noise of the experiment, the microwave pulses used to rotate the X and Y components to the Z axis for readout suffer from small systematic errors in their relative phase, which creates an offset from the proper rotation axis, and also in their duration, which creates an offset from the proper rotation length. The unitary operators above thus deviate from the ideal case and can be redefined, following Dobrovitski \emph{et al.} \cite{Dobrovitski2010}, by
\begin{align*}
U_{\mathrm{X}_{\frac{\pi}{2}}} &= \mathrm{exp} \left( -i \left(\vec{n}_{\mathrm{X}} \cdot \vec{\sigma}\right) \left( \frac{\pi}{2} + 2 \phi\right) \right)\\
U_{\mathrm{Y}_{-\!\frac{\pi}{2}}} &= \mathrm{exp} \left( -i \left(\vec{n}_{\mathrm{Y}} \cdot \vec{\sigma}\right) \left( -\frac{\pi}{2} + 2 \theta\right) \right),
\end{align*}
where $\vec{n}_{\mathrm{X}} = \left(1, \epsilon_y, \epsilon_z\right)/\left(1 + \epsilon_y^2 + \epsilon_z^2\right)^{1/2}$, $\vec{n}_{\mathrm{Y}} = \left( v_x, 1, v_z \right)/\left( 1 + v_x^2 + v_z^2\right)^{1/2}$, and $\vec{\sigma}$ is a vector composed of the Pauli matrices. Here, the $\epsilon$ and $v$ terms are the axial offsets and $\phi$ and $\theta$ are the length offsets, all of which are measured in units of angle. From calibration of the IQ modulator and the discrete nature of the delay generator that governs the length of the pulses, we estimate that these factors are at most $5^{\circ}$ in angular error and somewhat conservatively set the prior densities for each of these terms as a normal density of mean zero and standard deviation of $5^{\circ}$. The effect of this correction is most noticeable from the fact that the X and Y axis projection estimates almost always have larger uncertainties than the Z axis estimate.

Lastly, we use the non-informative reference prior\cite{Slater1998} for the density matrix,
\begin{equation}
\mathrm{prob}\left(\hat{\rho}\right) = 0.00513299\left(1 - r^2\right)^{-1/2} \left( \log \left[ \frac{\left( 1 - r \right)}{ \left( 1 + r \right)} \right] \right)^2 \sin \theta,
\end{equation}
where $r$ and $\theta$ are the standard spherical coordinates for the Bloch vector used to parameterize $\hat{\rho}$. As expected, given the large quantity of data collected, the choice of prior had no discernable effect on our inferences.

To obtain the marginal densities used for the point estimates and error bars of the projections and fidelities found in the main text, we use $\mathrm{MT\!-\!DREAM}_{\mathrm{ZS}}$, a Markov Chain Monte Carlo technique, to sample from the 11-dimensional posterior probability density \cite{Laloy2012}. This technique uses multiple random walk chains of a multiple-try Metropolis-Hastings rule \cite{Liu2000} applied to an adaptive proposal distribution generated from past samples to generate new samples in accordance with the posterior probability density. We found the chains to have fast convergence and good mixing properties, and the corresponding sampler output to have low autocorrelation using four independent chains and a multiple-try parameter of seven. The fidelities for experimental data are obtained by calculating the fidelity between the Bloch vectors from the random walk sampler and an ideal vector of unit length pointing along the same axis as the corresponding random walk sample. In this way, the fidelity is computed with respect to a perfect rotation of the state generated from the resonant laser pumping scheme described above. The point estimates of the projections and fidelities are the mean of the respective marginal densities while the error bars are the highest posterior density 68.2\% credible intervals \cite{Chen1999}.

In Figs.~\ref{fig:AzmInit},~\ref{fig:PolInit}, and~\ref{fig:GCInit}, we present the corresponding X, Y, and Z projections for the rotating-frame Bloch spheres describing arbitrary initialization via CPT presented in the main text, as well as some additional sets of projections and Bloch spheres not presented in the main text.
As described in the main text, all of these qubit Bloch spheres are within the $\omega_{mw} = \delta_{GS} / \hbar$ rotating frame.  This is because our standard qubit states $|0_g\rangle$ and $|\!+\!1_g\rangle$ are separated in energy by $\delta_{GS}$, and so each spin state precesses at the Larmor precession frequency of $\delta_{GS} / \hbar$. By reading out the  X, Y, and Z projections with ESR pulses phase-matched relative to the same $\omega_{mw} = \delta_{GS} / \hbar$ angular frequency, we capture a snapshot of a fixed point on this rotating-frame sphere. Conversely, in a system where the qubit states are not energy split, the Bloch sphere would not be precessing.  A CPT initialization laser pulse length of \unit{200}{\nano\second} is used for Figs.~\ref{fig:AzmInit},~\ref{fig:PolInit}, and~\ref{fig:GCInit}.  In some instances, noted below, this is not long enough to fully polarize the spin.

\begin{figure}[h!]
\begin{center}
\includegraphics{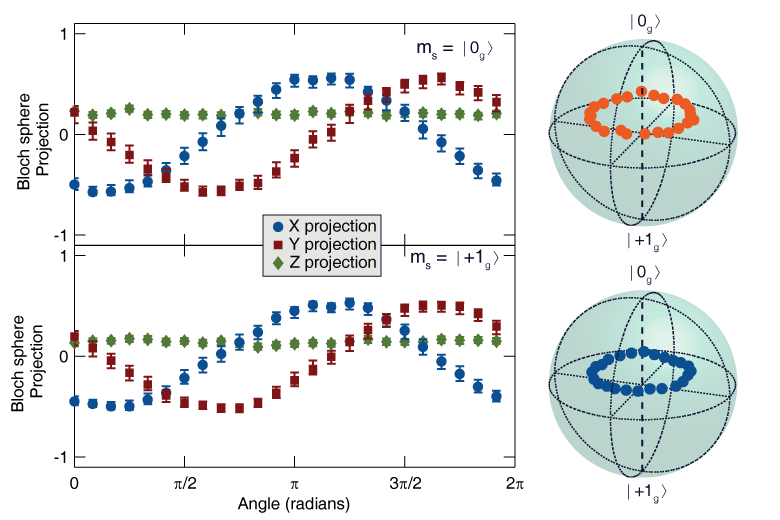}
  \caption[]{\label{fig:AzmInit}
\textbf{Projections for azimuthal initialization of spins.} The azimuthal location of the polarized spin state is rotated along the equatorial plane by varying the relative phase between the two colors.  X, Y, and Z projections are plotted on the left, and reconstructed Bloch spheres are plotted on the right. Top: Prior to the CPT interaction, the state was $\sf{|0_{g}\rangle}$.  Bottom: Prior to the CPT interaction, the state was in $\sf{|\!+\!1_{g}\rangle}$; this data is presented in Fig. 3A in Bloch sphere form.  Error bars on projections are the 68.2\% highest posterior density credible intervals from the Bayesian analysis.  }
\end{center}
\end{figure}

\begin{figure}[h!]
\begin{center}
\includegraphics{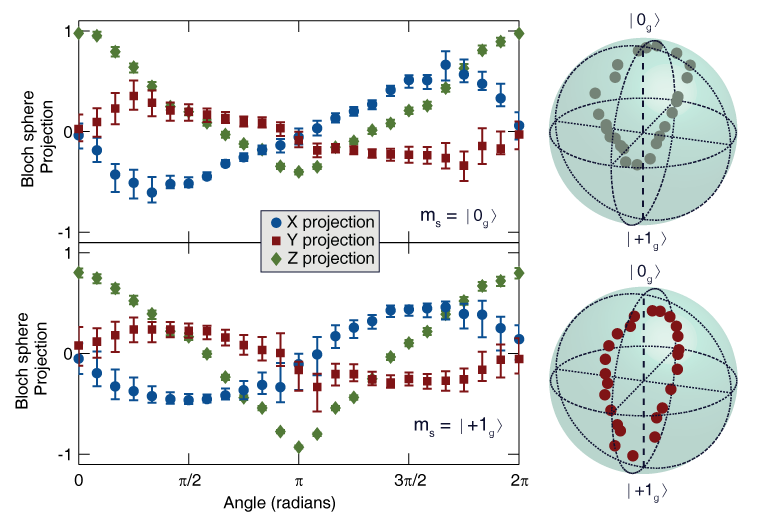}
  \caption[]{\label{fig:PolInit}
\textbf{Projections for polar initialization of spins.} We vary the polar location of the polarized spin state by varying the relative amplitude between the two colors.  X, Y, and Z projections are plotted on the left, while the same points in a Bloch sphere representation are plotted on the right. Top: Prior to the CPT interaction, the state was in $\sf{|0_{g}\rangle}$.  Bottom: Prior to the CPT interaction, the state was in $\sf{|\!+\!1_{g}\rangle}$; these data are also presented in Fig. 2B in Bloch sphere form.  Error bars on projections are the 68.2\% highest posterior density credible intervals from the Bayesian analysis.}
\end{center}
\end{figure}

Note that in Fig. \ref{fig:PolInit}, there is an asymmetry between the Z projection of the dark state near the poles depending on whether the state was prepared in $|0_{g}\rangle$ or $|\!+\!1_{g}\rangle$. This is partly due to the fact that a longer initialization pulse would be required to fully move the state to the opposite pole. In addition, some of this imbalance could also be due to differential spin coupling of the ISC and its resultant decay before readout occurs. This effect can also be seen in the Fig. \ref{fig:GCInit} data, where we combine phase and amplitude control of the light fields in order to initialize points along a great circle rotated $\pi/4$ off the equator.

\begin{figure}[h!]
\begin{center}
\includegraphics{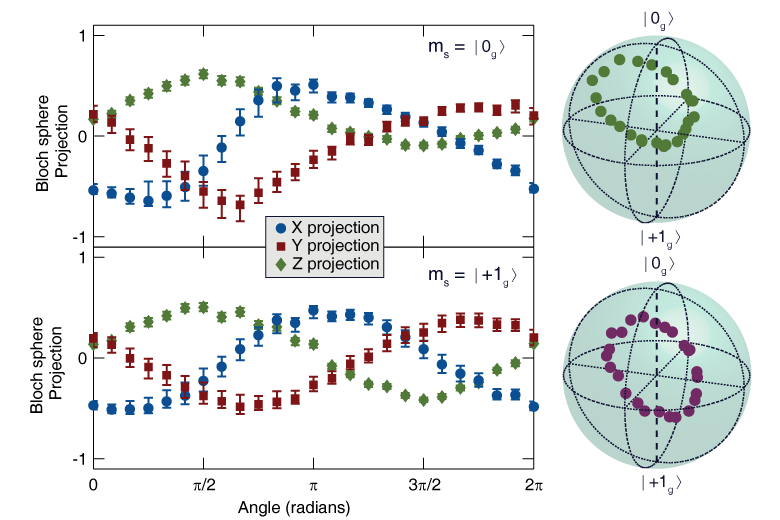}
  \caption[]{\label{fig:GCInit}
\textbf{Projections for initialization of spins along an off-axis great circle.} Here we vary both the relative amplitude and phase between the two colors to place the spins at points along a great circle, tilted  $\pi\sf{/4}$ off of the equator. Top: Prior to the CPT interaction, the state was in $\sf{|0_{g}\rangle}$.  Bottom: Prior to the CPT interaction, the state was in $\sf{|\!+\!1_{g}\rangle}$, this data is presented in Fig. 3D.  Error bars on projections are the 68.2\% highest posterior density credible intervals from the Bayesian analysis. }
\end{center}
\end{figure}

Finally, we should also note that the collection of Bloch sphere representations in all Main Text and Supplementary figures are not all viewed from the same vantage point, some were rotated to better show the results.   However, within a single figure the angle of view is fixed. One angle of view is used for the Bloch spheres in Fig. 2 and Fig. \ref{fig:InitModel} (Time dynamics of CPT initialization).  A second angle of view is used for the Bloch spheres in Fig. 3, Fig. \ref{fig:AzmInit}, \ref{fig:PolInit}, and \ref{fig:GCInit} (Arbitrary spin-state initialization).  A third angle of view is used for the Bloch spheres in Fig. 5, Fig. \ref{fig:XRot}, \ref{fig:YRot}, and \ref{fig:ZRot} (All-optical control of the NV center spin).

\clearpage

\vspace{24pt}
\section{Model of arbitrary initialization and rotation}
\vspace{12pt}


The Hamiltonian describing our system (Equation [1] from \textbf{Methods}) is presented below in matrix form,
$$
H =
 \begin{pmatrix}
  \Delta_L & 0 & \Omega \cos(\theta/2) & \Omega \cos(\theta/2) & 0 \\
  0 & \Delta_L & \Omega \sin(\theta/2)e^{i\phi}  & -\Omega \sin(\theta/2)e^{i\phi}   & 0 \\
  \Omega \cos(\theta/2) & \Omega \sin(\theta/2)e^{-i\phi}  & 0  & 0  & 0 \\
  \Omega \cos(\theta/2) & -\Omega \sin(\theta/2)e^{-i\phi} & 0  & -\delta_{e1}  & 0 \\
  0 & 0 & 0 & 0 & \epsilon_S
 \end{pmatrix}
$$
where the ordering of the states in the matrix is: $\{|+1_g\rangle,|0_g\rangle,|R_{e1}\rangle,|L_{e1}\rangle, |S\rangle\}$, $\Delta_L$ is the detuning of the laser frequency ($\omega_L$)  from resonance to the $|R_{e1}\rangle$ $\Lambda$ system, $\delta_{e1}$ is the separation of the excited state levels, $\Omega$ is the optical Rabi frequency, $\phi$ is the relative phase between the two coherent light fields, and $\tan(\theta/2)$ is the relative amplitude between the driving fields.  As such, $\phi$ and $\theta$ will describe the azimuthal and polar angle, respectively, of the resultant dark state.

The time evolution of the system is described by the Lindblad master
equation \cite{Lindblad1976,Breuer2002},
\begin{equation}
\dot\rho = i\left[\rho,H\right]
+ \sum_{\alpha, \alpha'} \Gamma_{\alpha \alpha'}  \left(\sigma_{\alpha'\alpha}\rho \sigma_{\alpha\alpha'}
                 -\frac{1}{2}\sigma_{\alpha\alpha}\rho  -\frac{1}{2}\rho\sigma_{\alpha\alpha} \right)
\equiv W\rho,
\end{equation}
with the Lindblad operators ${\sigma_{\alpha\alpha} = |\alpha\rangle\langle\alpha|} = \sigma_{\alpha'\alpha}^\dagger \sigma_{\alpha'\alpha}$ and ${\sigma_{\alpha'\alpha} = \sigma_{\alpha\alpha'}^\dagger = |\alpha'\rangle\langle\alpha|}$.
%
For $n=5$ levels, the density matrix $\rho$ is a Hermitian 5x5 matrix and can thus be
described by $n^2=25$ real parameters ($n^2-1=24$ including the normalization condition
$\mathrm{Tr}\left(\rho\right) =1$).  The superoperator $W$ can thus be viewed as a 25x25 matrix
with rank 24.
We denote the decay rate from the excited states (${E=L,R}$) to the
ground states ($G=0,1$) with ${\Gamma = \Gamma_{E_{e1},G_g}}$, the rate for inter-system
crossing from the excited states to the singlet ${\Gamma_i = \Gamma_{E_{e1},S}}$, and the
inverse intersystem crossing rate from $|S\rangle$ to one of the ground state levels
as ${\Gamma_i' = \Gamma_{S,G_{g}}}$. The spin relaxation rate in the ground state is
${\Gamma_1 =1/T_1= \Gamma_{+1_g,0_g}}$ and at sufficiently low temperature ${\Gamma_{0_g,+1_g}\approx 0}$.
The pure dephasing between the two ground state levels is denoted ${\gamma = 1/T_2 =\Gamma_{0_g,0_g}}$.
All other rates are set to zero.

The state of the system after optical excitation during time $t$ is obtained as
\begin{equation}
\rho(t) = e^{W t}\rho(0),\label{dynamics}
\end{equation}
where we choose one of the ground states as the initial state,
${\rho(0) = |0_g\rangle\langle0_g|}$ or ${\rho(0) = |+1_g\rangle\langle+1_g|}$.
We typically determine $\rho(t)$ by performing the exponentiation Eq.~(\ref{dynamics}) numerically.
The resulting Bloch vector in the ground state subspace can be obtained from
\begin{equation}
\mathbf{b}(t) = \mathrm{Tr} \left(\boldsymbol{\sigma}\rho(t)\right),
\end{equation}
where the components of $\boldsymbol{\sigma}$ are the Pauli matrices
in the ground-state subspace,
\begin{eqnarray}
\sigma_\mathrm{x} &=& |+1_g\rangle\langle 0_g| + |0_g\rangle\langle +1_g|, \\
\sigma_\mathrm{y} &=&  i (|+1_g\rangle\langle 0_g| - |0_g\rangle\langle +1_g|), \\
\sigma_\mathrm{z} &=& |0_g\rangle\langle 0_g| - |+1_g\rangle\langle +1_g|.
\end{eqnarray}

In the idealized case ${\Gamma_1 = \gamma = \Gamma_i =0}$, and with only one of the excited
levels included, the stationary state $\bar \rho$
in the long-time limit $t\gg 1/\Gamma$ obtained from ${\dot\rho=0}$ as the eigenvector of $W$
with eigenvalue $0$ is the dark state:
\begin{equation}
|D\rangle = \cos(\theta/2) |0_g\rangle -\exp(\mp i\phi)\sin(\theta/2) |+1_g\rangle
\end{equation}
where the upper (lower) sign holds for the single excited state level being ${E=R}$ (${E=L}$).
The fidelity of the state after a finite pumping time $t$ with realistic parameters
is then
\begin{equation}
F(t) = \langle D| \rho(t) |D\rangle.
\end{equation}
The experimentally obtained fidelity is shown in Fig. 2C in the main text.

We use this model to simulate the time evolution of the Bloch vector $\mathbf{b}(t)$ during \unit{500}{\nano\second} of CPT initialization plotted in Fig. 2B in the main text (Fig. \ref{fig:InitModel}) and the \unit{200}{\nano\second} of SRT rotation.  We fix the excited state splitting ${\delta_{e1} = \unit{180}{\mega\hertz}}$, and $\Gamma_i'$ using the known lower singlet lifetime of $371\,\mathrm{ns}$\cite{Robledo2011a}. We also fix the rate for both the $|R_{e1}\rangle$ and $|L_{e1}\rangle$ states into the singlet, $\Gamma_i$, as one half of the known ISC rate of $m_s = \pm 1$ spin states to the singlet of $\approx 74\,\mathrm{MHz}$, since $|R_{e1}\rangle$ and $|L_{e1}\rangle$ are composed of equal mixtures of $m_s = 0$ and $m_s = +1$ states. The detuning of the optical fields relative to $|R_{e1}\rangle$, $\Delta_L$, is fixed to the experimentally measured detuning for each operation (CPT initialization, SRT $\sigma_x$ rotation, SRT $\sigma_y$ rotation, SRT $\sigma_z$ rotation). Using a weighted least squares approach, we fit the overall driving amplitude, $\Omega$, relative amplitude, $\tan(\theta /2)$, relative phase of the driving fields, $\phi$, and the decay rates $\Gamma$, $\Gamma_1$, and $\gamma$, described above. All simulations show qualitative agreement with the experiment, however certain traces appear to have out-of-phase behavior of individual projections, marginal agreement with the initial state, or other disagreements. It appears that the model captures the essential physics but cannot fully account for certain ill-defined nuances such as the transients during the turn-on/off of the optical fields and effects related to the hyperfine spectrum and $T_2^*$. Therefore, the fitted values for the decay parameters are skewed by effects not considered in the model. A full set of fixed and fit parameters is found in Table \ref{tab:SimParams}.

\begin{figure}[]
\begin{center}
\includegraphics{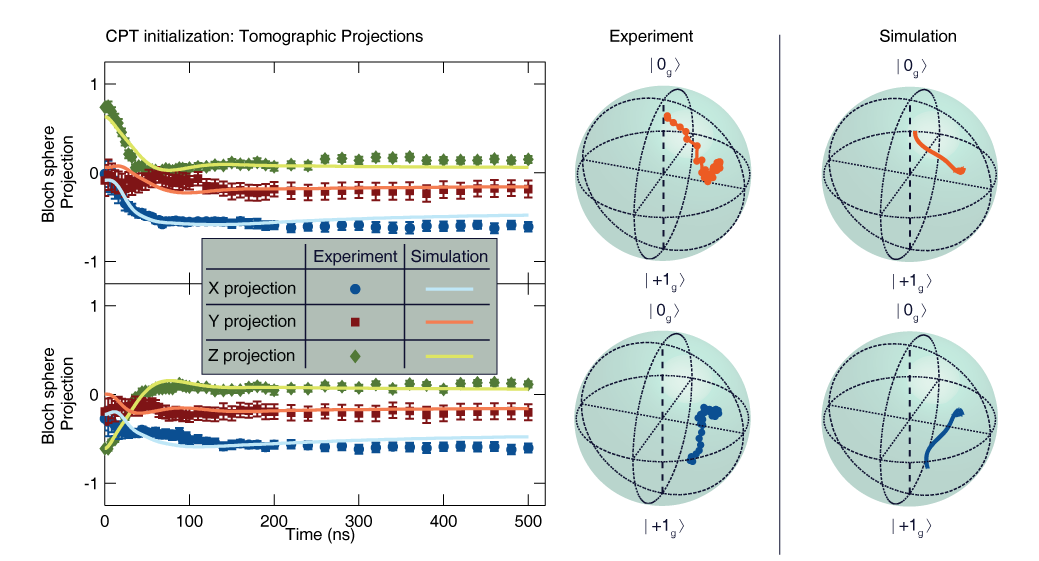}
  \caption[]{\label{fig:InitModel}
\textbf{Time dynamics of arbitrary initialization: theory vs. experiment.} As a function of CPT initialization pulse duration, tomographic reconstructions of the spin state are plotted alongside a simulation of the resultant state using the model. Top: Prior to the CPT interaction, the state was $\sf{|0_{g}\rangle}$.  Bottom: Prior to the CPT interaction, the state was in $\sf{|+1_{g}\rangle}$.  Error bars on projections are the 68.2\% highest posterior density credible intervals from the Bayesian analysis. Parameters to simulate spin-state initialization are found in Table \ref{tab:SimParams}. }
\end{center}
\end{figure}

As mentioned in the main text, it may be possible to rotate about any arbitrary axis, but there are a few considerations to be made.  SRT are a dispersive interaction whose strength is proportional to $1/\Delta_L$, whereas CPT is an absorptive process whose interaction strength is proportional to $1/\Delta_L^2$.  To take advantage of SRT, sufficient detuning is necessary to diminish absorptive effects that are non-unitary and cause the spin to polarize along the dark state rather than rotate.  In the case of rotations about an equatorial axis, we take advantage of the two competing $\Lambda$ systems by tuning $\omega_L$ exactly between $|R_{e1}\rangle$ and $|L_{e1}\rangle$ resonances.  These $\Lambda$ systems have opposite equatorial bright states and we detune from both in opposite directions, essentially causing SRT effects to add constructively and the CPT effects to add destructively.  In order to rotate about an axis off of the equator, care must be taken to couple more strongly to one of the $\Lambda$ systems than the other because their corresponding bright states are no longer orthogonal, as the bright states for each $\Lambda$ system will have orthogonal azimuthal phases but the same polar component. As discussed above, a more widely spaced anticrossing to decrease the competition between the two upper states would aid in ensuring unitary non-equatorial axial rotations.  The special case of rotation about the polar axis, which we demonstrate in Fig. 5B, was also shown in Buckley \textit{et al.}\cite{Buckley2010} but not in the context of a $\Lambda$ system.  In our case, we sufficiently detune $\sim$\unit{450}{\mega\hertz} from the $|R_{e1}\rangle$ such that any effects from CPT are greatly diminished.

\begin{figure}[]
\begin{center}
\includegraphics{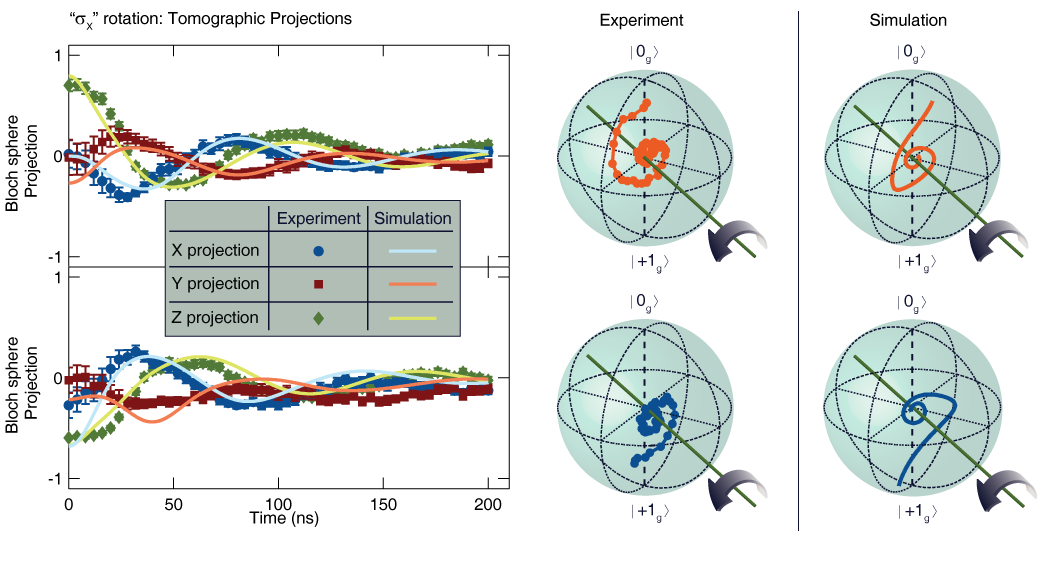}
  \caption[]{\label{fig:XRot}
\textbf{Time dynamics of arbitrary coherent $\sigma_{\sf{X}}$ rotation: theory vs. experiment.} As a function of the duration of a SRT $\sigma_{\sf{{X}}}$ rotation pulse duration, projections of the resultant spin state are plotted alongside a simulation of the resultant state using the model. Top: Prior to the SRT interaction, the state was $\sf{|0_{g}\rangle}$.  Bottom: Prior to the SRT interaction, the state was in $\sf{|+1_{g}\rangle}$.  Error bars on projections are the 68.2\% highest posterior density credible intervals from the Bayesian analysis. Parameters to simulate the $\sigma_{\sf{X}}$ rotation are found in Table \ref{tab:SimParams}.}
\end{center}
\end{figure}

\begin{figure}[]
\begin{center}
\includegraphics{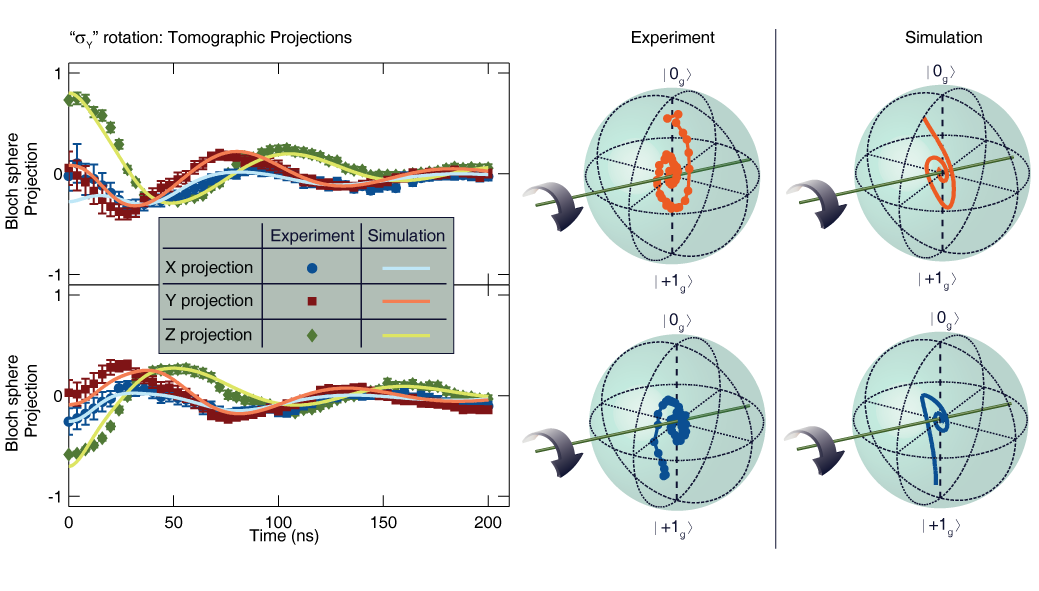}
  \caption[]{\label{fig:YRot}
\textbf{Time dynamics of arbitrary coherent $\sigma_{\sf{Y}}$ rotation: theory vs. experiment.} As a function of the duration of a SRT $\sigma_{\sf{Y}}$ rotation pulse duration, projections of the resultant spin state are plotted alongside a simulation of the resultant state using the model. Top: Prior to the SRT interaction, the state was $\sf{|0_{g}\rangle}$.  Bottom: Prior to the SRT interaction, the state was in $\sf{|\!+\!1_{g}\rangle}$.  Error bars on projections are the 68.2\% highest posterior density credible intervals from the Bayesian analysis. Parameters to simulate the $\sigma_{\sf{Y}}$ rotation are found in Table \ref{tab:SimParams}.}
\end{center}
\end{figure}

\begin{figure}[]
\begin{center}
\includegraphics{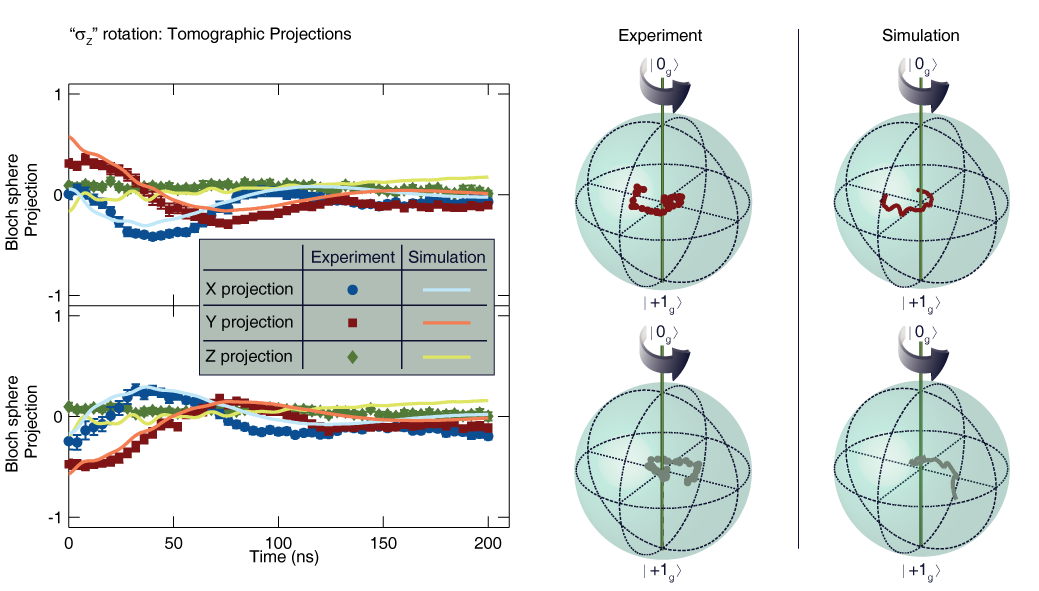}
  \caption[]{\label{fig:ZRot}
\textbf{Time dynamics of arbitrary coherent $\sigma_{\sf{Z}}$ rotation: theory vs. experiment.} As a function of the duration of a SRT $\sigma_{\sf{Z}}$ rotation pulse duration, projections of the resultant spin state are plotted alongside a simulation of the resultant state using the model. Top: Prior to the SRT interaction, the state was $\sf{|X_{g}\rangle}$.  Bottom: Prior to the SRT interaction, the state was in $\sf{|-X_{g}\rangle}$.  Error bars on projections are the 68.2\% highest posterior density credible intervals from the Bayesian analysis. Parameters to simulate the $\sigma_{\sf{Z}}$ rotation are found in Table \ref{tab:SimParams}.}
\end{center}
\end{figure}

\begin{table}
\begin{tabular}{c||c||c|c|c|c}
                        & Parameter     & CPT initialization &  $\sigma_{\mathrm{X}}$ & $\sigma_{\mathrm{Y}}$ & $\sigma_{\mathrm{Z}}$\\
\hline\hline
Fixed Level   & $\delta_{e1}$ &\unit{180}{\mega\hertz}&\unit{180}{\mega\hertz} &\unit{180}{\mega\hertz}   &\unit{180}{\mega\hertz} \\
Parameters    & $\Delta_{L}$  &\unit{-0.684}{\mega\hertz}&\unit{-90}{\mega\hertz}&\unit{-90}{\mega\hertz}   &\unit{-450}{\mega\hertz} \\
\hline
Optical  & $\Omega$&\unit{46.507}{\mega\hertz}&\unit{62.021}{\mega\hertz}&\unit{62.756}{\mega\hertz}&\unit{84.104}{\mega\hertz} \\
Driving Field                        & $\theta$      &\unit{1.708}{\radian}&\unit{4.774}{\radian}&\unit{1.763}{\radian}     & $\pi$ rad \\
Parameters                        & $\phi$        &\unit{0.395}{\radian}&\unit{4.152}{\radian}&\unit{2.683}{\radian}     & \unit{0.424}{\radian} \\
\hline
Decay         & $\Gamma$   &\unit{35.114}{\mega\hertz}&\unit{17.115}{\mega\hertz}&\unit{19.719}{\mega\hertz}    & 0 \\
Parameters    & $\Gamma_i$ &\unit{37}{\mega\hertz}&\unit{37}{\mega\hertz}&\unit{37}{\mega\hertz}    & \unit{37}{\mega\hertz}\\
                        & $\Gamma_i'$&\unit{2.701}{\mega\hertz}&\unit{2.701}{\mega\hertz}&\unit{2.701}{\mega\hertz}       & \unit{2.701}{\mega\hertz} \\
                        & $\Gamma_1$ & \unit{0.373}{\mega\hertz}                   & 0               & 0            & 0 \\
                        & $\gamma$   & 0           & 0               & 0            & \unit{28.459}{\mega\hertz} \\
\hline
Initial State           & $r_A$      & 0.640                  &0.839               &0.852        &0.602\\
Parameters              & $\theta_A$ &\unit{0.164}{\radian}&\unit{0.327}{\radian}&\unit{0.347}{\radian}& \unit{1.844}{\radian}\\
                        & $\phi_A$   &\unit{2.526}{\radian}&\unit{4.705}{\radian}&\unit{2.850}{\radian}& \unit{1.471}{\radian}\\
                        & $r_B$      & 0.649             & 0.977               & 0.752                &0.621\\
                        & $\theta_B$  &\unit{2.788}{\radian}&\unit{2.325}{\radian}&\unit{2.774}{\radian}& \unit{1.870}{\radian}\\
                        & $\phi_B$    &\unit{3.122}{\radian}&\unit{3.450}{\radian}&\unit{3.472}{\radian}& \unit{4.425}{\radian}\\
\end{tabular}
\caption{\label{tab:SimParams} \textbf{Simulation Parameters.}  Simulation parameters for CPT initialization, $\sigma_{\sf{X}}$, $\sigma_{\sf{Y}}$, and $\sigma_{\sf{Z}}$.  In the case of CPT initialization, $\sigma_{\sf{X}}$, and $\sigma_{\sf{Y}}$, the initial state parameters state $\sf{A}$ refers to $\sf{|0_{g}\rangle}$ and state $\sf{B}$ refers to $\sf{|\!+\!1_{g}\rangle}$. For $\sigma_{\sf{Z}}$, state $\sf{A}$ corresponds to $\sf{|X_{g}\rangle}$ and state $\sf{B}$ corresponds to $\sf{|\!-\!X_{g}\rangle}$.
}
\end{table}

\clearpage
\vspace{24pt}
\section{Sources of decoherence}
\vspace{12pt}

The closeness in energy of the two anticrossed eigenstates, $|R_{e1}\rangle$ and $|L_{e1}\rangle$, causes a loss in fidelity.  This is because even when resonantly tuned to a single eigenstate, there is still off-resonant coupling to the other $\Lambda$ system. As the phases of the resultant dark states  between the two are orthogonal, the overall length of the vector pointing to the final state within the Bloch sphere is reduced.  The model shows good qualitative agreement to our presented data, revealing a similarly mixed state as a result of the competing $\Lambda$ systems. Therefore, this competition is one of the most significant source of decoherence in our measurement.  Finding a more widely spaced anticrossing would help alleviate this issue by decreasing the coupling to the other state; within the model, a purer final state (higher fidelity) results with increasing separation of the two excited states.  Experimentally, one method to achieving more widely spaced anticrossing would be to slightly misalign the field because off-axis fields will increase the separation in any anticrossing.  However, it should be noted that a misalignment of the field will also change the eigenstates, possibly producing more transitions from spin-mixing into the $|\!-\!1_g\rangle$ spin sublevel, which also causes a loss in fidelity.

Further sources of decoherence include the transverse inhomogeneous spin coherence time $T_2^* \sim$\unit{1}{\micro\second} as well as the fact that our spin sublevels are further split into three nuclear hyperfine states due to the $^{14}$N in our NV center, with each transition split by $\sim$\unit{2}{\mega\hertz}.  Since our measurements are averages over $\sim\!10^6$ replications of the same experiment, any individual replication has an equal probability of being in any of the three hyperfine states, causing our selected $\omega_{mw}$ to be $\sim\!\pm$\unit{2}{\mega\hertz} detuned from the actual ground state splitting two thirds of the time.  We note that no nuclear polarization was observed at this excited state anticrossing at cryogenic temperatures.  This hyperfine spectrum and finite $T_2^*$ effectively set limits on how long our $\Lambda$ interaction remains phase coherent with the spin, reducing transverse coherence of the interaction.  The resultant steady state is a balance between the strength of the interaction and these decoherence mechanisms (along with the other mechanisms mentioned in this section).  As such, an echo sequence would not eliminate these effects while the interaction is taking place, but only during the rest of the measurement sequence.  In addition, as the NV center is a solid-state defect, spectral diffusion causes a broadening of the natural linewidth of the resonances\cite{Fu2009}, which will also contribute to a lower fidelity.

Finally, since we are examining a $\Lambda$ system contained within a more complex level structure, the spin will end up outside of our qubit subspace a fraction of the time.  First of all, the spin can end up passing through the long-lived ISC, which is accounted for in the model as a singlet level, and does manifest as a decoherence mechanism. Secondly, since our qubit states, $|0_g\rangle$ and $|\!+\!1_g\rangle$, are a subspace of a spin-triplet ground state, the spin can transition to $|\!-\!1_g\rangle$ some of the time.  This is because, experimentally, the excited state levels $|R_{e1}\rangle$ and $|L_{e1}\rangle$ each contain anywhere from 1\% to 3\% of the excited state spin sublevel $|\!-\!1_{e1}\rangle$.  A spin in the$|\!-\!1_g\rangle$ sublevel is not resonant with our red lasers and will therefore appear dark.  This suggests that there might be higher fidelities resulting from implementation of this type of control within a system that lacks decay mechanisms such an intersystem crossing.

\begin{figure}[]
\begin{center}
\includegraphics{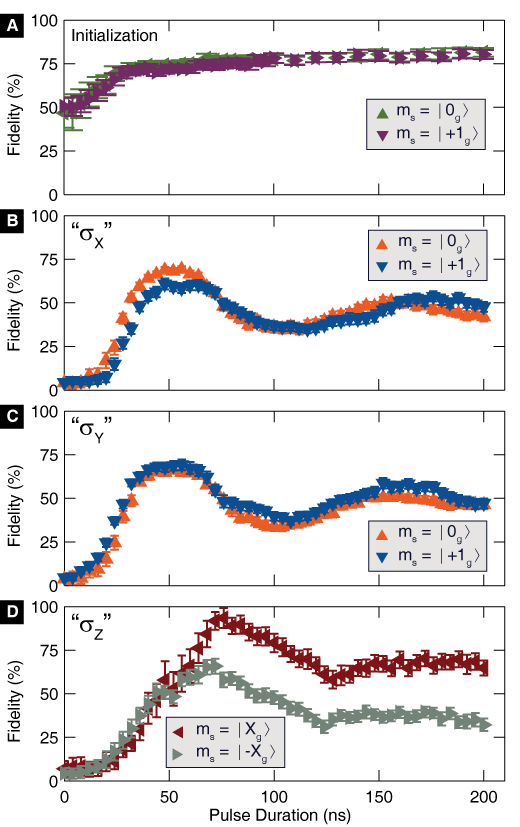}
  \caption[]{\label{fig:Fidelities}
\textbf{Fidelities of initialization and rotation.} $\textbf{A}$, Fidelity of initialization as a function of pulse duration for an initial state $\sf{|0_{g}\rangle}$ and $\sf{|\!+\!1_{g}\rangle}$. $\textbf{B}$, Process fidelity of $\sigma_{\sf{X}}$ rotation compared to a perfect $\pi$ rotation, as a function of pulse duration for an initial state $\sf{|0_{g}\rangle}$ and $\sf{|\!+\!1_{g}\rangle}$. $\textbf{C}$, Process fidelity of $\sigma_{\sf{Y}}$ rotation as compared to a perfect $\pi$ rotation, as a function of pulse duration for an initial state $\sf{|0_{g}\rangle}$ and $\sf{|\!+\!1_{g}\rangle}$. $\textbf{D}$, Process fidelity of $\sigma_{\sf{Z}}$ rotation as compared to a perfect $\pi$ rotation, as a function of pulse duration for an initial state $\sf{|X_{g}\rangle}$ and $\sf{|\!-\!X_{g}\rangle}$.  For all process fidelities, we compare the resultant state, as a function of pulse duration, to a state exactly $\pi$ out-of-phase with the initial state about the rotation axis. We renormalize the initial mixed state to a pure state in order to compute the fidelity loss from the CPT or SRT process alone.  }
\end{center}
\end{figure}

Through our tomographic reconstructions, we are able to determine that the fidelity of our spin-state initialization saturates at roughly 80\% after \unit{100}{\nano\second} (Fig. \ref{fig:Fidelities}A).  The fidelities of a  $\sigma_{\mathrm{X}}$ or $\sigma_{\mathrm{Y}}$  $\pi$ rotation are as high as 69\% if we assume a pure initial state as the initial state instead the mixed state used in the experiment (Fig. \ref{fig:Fidelities}B, C).  A $\sigma_{\mathrm{Z}}$ $\pi$ rotation could have process fidelities as high as 90\% (Fig. \ref{fig:Fidelities}D), but as the initial states used for that experiment were of lower fidelity than any other experiment, this estimation may be rather optimistic.

\vspace{24pt}
\section{Arbitrary spin-state readout}
\vspace{12pt}
The following pulse sequence was used to perform the DBP readout protocol.

\begin{figure}[h!]
\begin{center}
\includegraphics{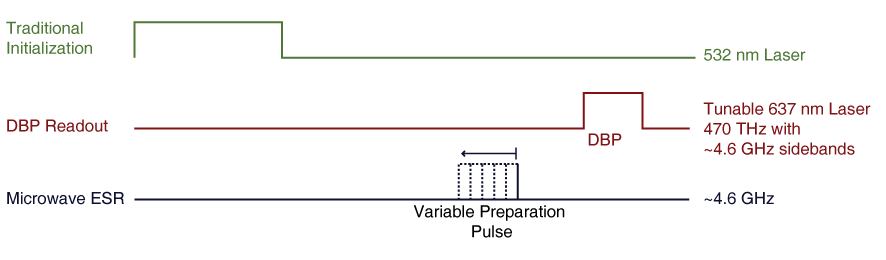}
  \caption[]{\label{fig:DBPPulse}
\textbf{Pulse sequence for arbitrary spin-state readout.} The above pulse sequence was used for the data presented in Main Text Fig. 4A, B.  The DBP readout experiments consisted of $\sf{3.75 \times 10^{6}}$ iterations of this pulse sequence.  }
\end{center}
\end{figure}

Our DBP optical spin readout protocol begins with preparation with non-resonant \unit{532}{\nano\meter} excitation to prepare the state in $|0_g\rangle$.  This is followed by an on-chip microwave pulse to prepare the spin at various points about the Bloch sphere.  In the case of Fig. 4A, the pulse corresponds to a $\pi/2$ pulse, and its phase is varied to place the spin at various points about the Bloch sphere equator.  In the case of Fig. 4B, its pulse duration is varied to induce Rabi oscillations between the ground state spin sublevels, $|0_g\rangle$ and $|\!+\!1_g\rangle$, initializing the spin at various points along a meridian.  A DBP pulse of \unit{400}{\nano\second} is used to read out the spin state and is delayed \unit{450}{\nano\second} from the microwave initialization pulse, much like the delay between the CPT initialization and the ESR  $\pi/2$ projection pulses for quantum state tomography of CPT and SRT. Spin readout protocols such as QST could be performed directly with DBP mitigating the need for ESR; however, the differing PL contrasts between polar and azimuthal spin readout as a result of qubit dephasing must be appropriately calibrated out (Fig. 4A,B).


To determine the quality of DBP spin readout vs. traditional ``green'' spin readout via the ISC, we analyze the signal-to-shot-noise ratio of both readout methods.  Shot noise of counted photons is the primary noise source in our data, which is a common feature for these types of experiments with proper mitigation of any systematic errors such as experimental drift.  Each data point in Fig. 4A,B consists of summed photon counts of $n = 3.75 \times 10^6$ individual spin readouts using a \unit{400}{\nano\second} DBP pulse.  When fit to a sinusoid, the polar Bloch sphere readout in Fig. 4B results in $I_{BZ} = 4,\!850$ counts if the spin is in the bright state and $I_{DZ} = 1,\!750$ counts if the spin is in the dark state (averaged fit values of $|0_g\rangle$ and $|\!+\!1_g\rangle$ data).  The equatorial Bloch sphere readout in Fig. 4B with reduced contrast primarily due to dephasing results in $I_{BX} = 5,\!380$ counts if the spin is in the bright state and $I_{DX} = 3,\!160$ counts if the spin is in the dark state (averaged fit values of $|X_g\rangle$ and $|\!-\!X_g\rangle$ data).  To compare to traditional ISC ``green'' readout, we will assume reasonable numbers for green readout with our specific confocal setup: a \unit{400}{\nano\second} readout window with $20,\!000$ Cts/s for the $m_s = 0$ spin state, and a 30\% reduction of counts for $m_s = \pm1$ spin states during this window.  For an equivalent averaging time ($n=3.75 \times 10^6$), this corresponds to green spin readout counts of: $I_{BG} = 30,\!000$ and $I_{DG} = 21,\!000$.  The number of spin readouts ($N_i$) required for the noise standard deviation ($\sqrt{n}$) to be roughly equal to the full-scale readout contrast is:
\begin{equation}
N_i=\frac{n/2 \times \left(I_{Bi}+I_{Di}\right)}{\left(I_{Bi}-I_{Di} \right)^2 }
\end{equation}
the number of individual spin readouts to get a signal-to-noise ratio of unity for these three readout techniques (Polar DBP ($N_{\mathrm{Z}}$), Azimuthal DBP ($N_{\mathrm{X}}$), and green ISC ($N_{\mathrm{G}}$) are respectively:
\vspace{6pt}
\begin{center}\hspace{20cm} $N_{\mathrm{Z}}=1,\!290$, $N_{\mathrm{X}}=3,\!250$, $N_{\mathrm{G}}=1,\!180$\end{center}
\vspace{6pt}
Therefore, using these rough numbers, DBP on a polar axis (``spin up'' vs. ``spin down'') has a signal-to-noise ratio comparable to green spin up/down readout, while DBP on a precessing equatorial axis (``spin left'' vs. ``spin right'') takes roughly three times as much averaging to get the same signal-to-noise ratio as the other two techniques.

\clearpage
\vspace{24pt}
\section{All-optical Ramsey measurement}
\vspace{12pt}
We use the pulse sequence in Fig. \ref{fig:RamseySeq} to take the data for the top portion of Fig. 4C and in Fig. \ref{fig:MoreOpticalRamsey}.
\begin{figure}[h!]
\begin{center}
\includegraphics{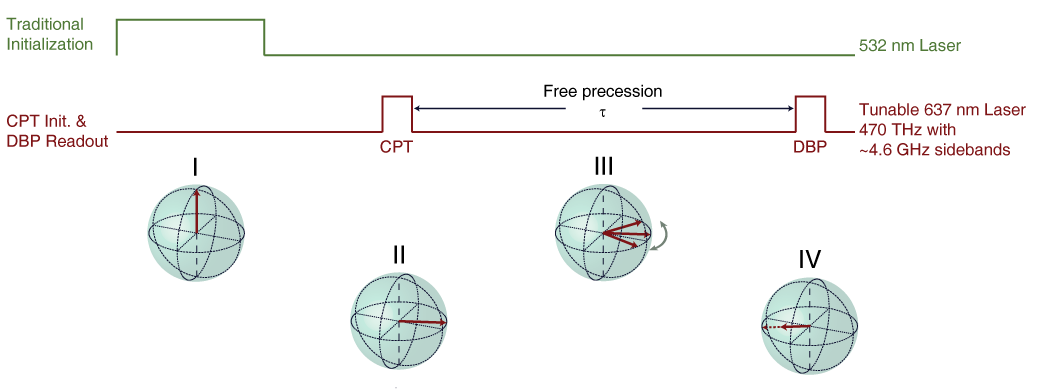}
  \caption[]{\label{fig:RamseySeq}
\textbf{Pulse sequence for arbitrary spin-state readout.} I. We begin with a non-resonant 532\nobreakspace nm excitation to prepare the state in $\sf{|0_{g}\rangle}$.  II. The state is then initialized onto the equator with a CPT pulse of length 50\nobreakspace ns, in one of four azimuthal positions $\pi \sf{/2}$ out-of-phase from one another other.  III. The spin state dephases during free precession, $\tau$, which is varied.  IV. The state is then read out with a DBP pulse of length 50\nobreakspace ns of a fixed phase.  The data of two opposite CPT pulse phases were subtracted and plotted in the top of Fig. 4C. The EOM microwave detuning,($\omega_{\sf{mw}}\sf{/2}$$\pi-\delta_{\sf{GS}}/h$), is $\sf{\sim}$7.5\nobreakspace MHz from the mean qubit precession.}
\end{center}
\end{figure}

\begin{figure}[h!]
\begin{center}
\includegraphics{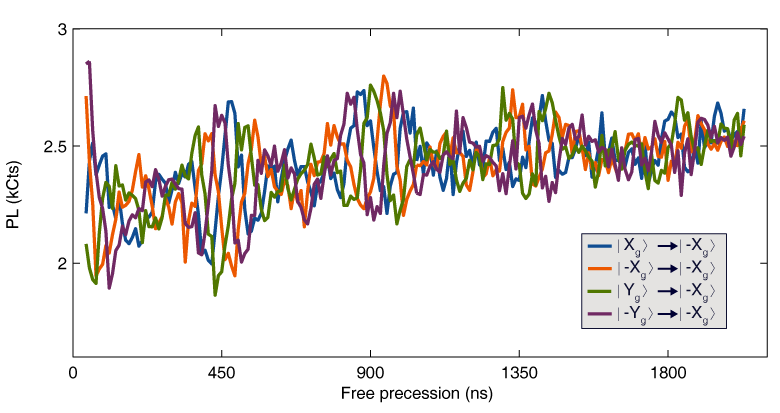}
  \caption[]{\label{fig:MoreOpticalRamsey}
\textbf{Non-subtracted all-optical Ramsey measurement.} Four different phases of a time-domain Ramsey experiment were measured and are plotted above.  The overall background is due to the CPT pulse populating the dark ISC, which subsequently decays over time, making the DBP PL brighter as it gets further away from the CPT pulse. }
\end{center}
\end{figure}

The Ramsey data plotted in Fig.~4C of the main text shows the difference in
measured PL for two orthogonal initial spin projections, $\Delta\mathrm{PL} =
\mathrm{PL}(\left|-X_\mathrm{g}\right\rangle) -
\mathrm{PL}(\left|X_\mathrm{g}\right\rangle)$.  The difference cancels the
contribution to the PL from the ISC decay in the optical Ramsey measurements
(Fig. \ref{fig:MoreOpticalRamsey}), and effectively measures the projection $\langle S_X\rangle$
of the spin at the end of the free precession period.  In both cases the data
is fit to a function of the form
\begin{multline}\label{eq:RamseyFit}
  \Delta\mathrm{PL} =
  A\exp\left(-\frac{\tau^2}{2T_2^{\ast2}}\right)\Bigl\lbrace C_1\cos\bigl[(\delta
  \omega-\omega_\mathrm{HF})(\tau-\tau_0)\bigr] \\ +
  \cos\bigl[\delta
  \omega(\tau-\tau_0)\bigr]
  +C_2\cos\bigl[(\delta
  \omega+\omega_\mathrm{HF})(\tau-\tau_0)\bigr]\Bigr\rbrace,
\end{multline}
where
\begin{equation}
    \delta\omega = \omega_{mw}-\delta_{GS}/\hbar,
\end{equation}
and includes the threefold hyperfine coupling to the $^{14}$N nuclear spin
with frequency, $\omega_\mathrm{HF}$, inhomogeneous dephasing with
characteristic time $T_2^\ast$, and independent amplitudes for the three
hyperfine components. The temporal offset $\tau_0$ accounts for the effects
of finite-duration initialization and readout pulses. Best-fit parameter
values for the curves plotted in Fig.~4C are provided in
Table~\ref{tab:RamseyFits}.

\begin{table}
\begin{tabular}{@{}lccccccc@{}}
\hline
 & $T_2^\ast$ & $\delta\omega/2\pi$ & $\omega_\mathrm{HF}/2\pi$ & $\tau_0$ & $A$ & $C_1$ & $C_2$ \\
 & ($\mu$s) & (MHz) & (MHz) & (ns) & (Cts) & & \\
  \hline
  Optical & $1.13\pm0.05$ & $7.52\pm0.01$ & $2.19\pm0.01$ & $13\pm1$ & $253\pm13$ & $1.36\pm0.07$ & $0.64\pm0.05$ \\
  ESR & $1.01\pm0.03$ & $7.34\pm0.01$ & $2.20\pm0.01$ & $-12\pm2$ & $1230\pm40$ & $1.16\pm0.04$ & $0.68\pm0.04$ \\
  \hline
\end{tabular}
\caption{\label{tab:RamseyFits}
Best-fit parameter values from fits of the model of Eq.~\ref{eq:RamseyFit} to
the data in Fig.~4C of the main text.  Uncertainties are standard error.}
\end{table}

\clearpage

\vspace{24pt}
\section{All-optical Hahn echo measurement}
\vspace{12pt}
We use the pulse sequence in Fig. \ref{fig:OpticalHahnSequence} to take the data in Fig. 5C.
\begin{figure}[h!]
\begin{center}
\includegraphics{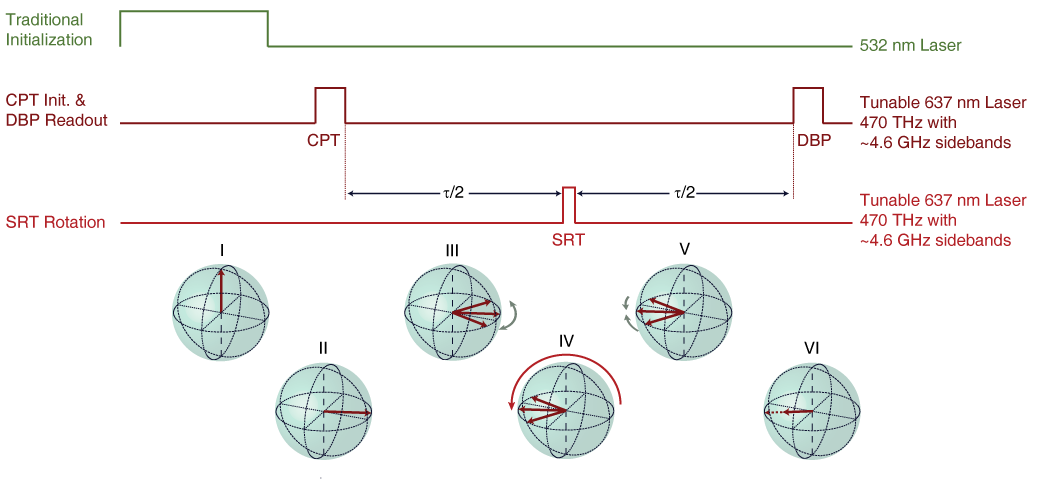}
  \caption[]{\label{fig:OpticalHahnSequence}
\textbf{Pulse sequence for all-optical Hahn echo measurement. } I. A 532\nobreakspace nm excitation is used to prepare the state into $\sf{|0_g\rangle}$.  II. The spin is initialized on the Bloch equator with a CPT pulse (Fig. 3A).  III. Dephasing of the spin state occurs during free precession for a delay  $\tau \sf{/2}$.  IV.  The spin is then rotated by a SRT   pulse (Fig. 5A).  V. Rephasing of the spin state occurs during a period of $\tau \sf{/2}$. VI. Finally the spin state is readout along the equator with a DBP pulse (Fig. 4A), corresponding to a bright state either in-phase or $\pi$ out-of-phase with the CPT pulse.  The Hahn precession time, $\tau$, is varied.}
\end{center}
\end{figure}

In Fig. 5C, the all-optical Hahn echo measurement is plotted on top of the data set for the ESR-based Hahn echo. We use least squares to fit the function ${\Delta \mathrm{PL} = A\exp\left( -\left(\tau / T_2\right)^{3} \right)}$ to our data and infer $T_2 \sim$\unit{900}{\micro\second} for both measurements.  The fitting parameters are provided in Table \ref{tab:HahnFits}.


%

\begin{table}[h!]
\begin{center}
\begin{tabular}{@{}lccccccc@{}}
\hline
            & $T_2$           & A \\
            & ($\mu$s)        & (Cts) \\
 \hline
 Optical    &   $893 \pm 51$  & $538 \pm 29$ \\
 ESR        &   $909 \pm 30$  & $2991 \pm 99$ \\
  \hline
\end{tabular}
\caption{\label{tab:HahnFits} Best-fit parameter values from fits of the
model to the data in Fig.~5D of the main text.
Uncertainties are standard error.}
\end{center}
\end{table}

%
%
%
%
%
%

\renewcommand{\refname}{Full References}

\end{document}